\documentclass[conference,compsoc]{IEEEtran}
\PassOptionsToPackage{hyphens}{url}

\usepackage[T1]{fontenc}
\usepackage[utf8]{inputenc}

\usepackage{booktabs}
\usepackage{listings}
\usepackage{mdframed}
\usepackage{graphicx}
\makeatletter
\def\input@path{{figures/}}
\makeatother
\usepackage{xspace}
\usepackage{amsmath}
\usepackage{xcolor}
\usepackage[normalem]{ulem} 
\usepackage{colortbl}
\usepackage{multirow}
\usepackage{tabularx}
\usepackage{pifont}

\usepackage{cite}
\usepackage{url}
\usepackage[hidelinks]{hyperref}  

\usepackage{arydshln} 
\usepackage{enumitem}

\usepackage{float}
\newfloat{listing}{htbp}{lol}
\floatname{listing}{Listing}

\newcommand{\malicious}{MalWild}
\newcommand{\malbench}{MalSkillBench}

\newcommand{\malbenchwild}{MalSkillBench-Wild}
\newcommand{\malbenchsyn}{MalSkillBench-Syn}
\newcommand{\trustbench}{SkillTrustBench}
\newcommand{\oasb}{OASB}
\newcommand{\skillsbench}{SkillBench-1650}
\newcommand{\skillinject}{Skill-Inject}

\newcommand{\maliciousshort}{MalWild}
\newcommand{\malbenchshort}{MSB}
\newcommand{\malbenchwildshort}{MSB-Wild}
\newcommand{\malbenchsynshort}{MSB-Syn}
\newcommand{\trustbenchshort}{STB}
\newcommand{\skillsbenchshort}{SkillBench}
\newcommand{\skillinjectshort}{Sk-Inject}
\newcommand{\skillsafetybench}{SkillSafetyBench}
\newcommand{\skillsafetybenchshort}{SSB}

\newcommand{\qwenfull}{Qwen3.6-35B-A3B}
\newcommand{\deepseekfull}{DeepSeek-V4-Flash}
\newcommand{\qwen}{Qwen3.6}
\newcommand{\deepseek}{DeepSeek-V4}
\newcommand{\gpt}{GPT-5.4}

\newcommand{\skillscanner}{skill-scanner}
\newcommand{\skillspector}{skillspector}
\newcommand{\skillscansecurity}{skillscan-security}
\newcommand{\skillvetter}{skill-vetter}
\newcommand{\aiinfraguard}{AI-Infra-Guard}
\newcommand{\clawmod}{ClawHub-Mod}
\newcommand{\snykagentscan}{Snyk Agent Scan}
\newcommand{\socketscanner}{Socket}
\newcommand{\genagenttrusthub}{Gen Agent Trust Hub}

\definecolor{bestgreen}{HTML}{0B4A22}
\definecolor{worstred}{HTML}{7A1119}
\newcommand{\best}[1]{\textcolor{bestgreen}{\textbf{#1}}}
\newcommand{\worst}[1]{\textcolor{worstred}{\textbf{#1}}}

\newcommand{\fpheat}[3]{\cellcolor[HTML]{#1}\textcolor{#2}{#3}}

\newif\ifdraftnotes
\draftnotestrue
\definecolor{juncomment}{HTML}{1A5FB4}

\makeatletter
\newcommand{\nosectpunct}{\def\@IEEEsectpunct{}}
\makeatother

\definecolor{sourcegreen}{HTML}{18723B}
\definecolor{editorange}{HTML}{B85B00}

\definecolor{deletered}{HTML}{BA2525}
\definecolor{newblue}{HTML}{1755B5}

\begin{document}

\title{Demystifying Detectors Guarding Agent Skill Hubs}


\author{%
\IEEEauthorblockN{%
Kang Yang\IEEEauthorrefmark{1},
Kevin Ying\IEEEauthorrefmark{2},
Xiaojing Liao\IEEEauthorrefmark{3},
Guanhong Tao\IEEEauthorrefmark{1},
Jun Xu\IEEEauthorrefmark{1},
}
\IEEEauthorblockA{%
\IEEEauthorrefmark{1}University of Utah, 
\IEEEauthorrefmark{2}Skyline High School,
\IEEEauthorrefmark{3}University of Illinois Urbana-Champaign 
}
}

\maketitle

\begin{abstract}
Agent Skills has gained widespread adoption as an open standard for extending agent capabilities. To facilitate skill sharing, public skill hubs have grown in popularity and increasingly rely on automated security detectors to identify malicious skills at scale. Yet, the effectiveness and limitations of these detectors remain insufficiently understood. 
To fill this gap, we present the first systematic study of four hub-adopted and two community-provided malicious skill detectors on 1147 agent skills from seven datasets. Our evaluation reveals that even the best-performing hub-adopted detector by overall $F_1$, \skillscanner{} with \qwen{}, still misses 4.9\% of malicious skills and incorrectly flags 17.8\% of benign skills.

To understand these failures, we manually analyze detectors’ findings and decision-making processes, identifying two recurring failure mechanisms. False positives often arise when detectors flag sensitive operations, such as credential access, without considering whether they serve the skill's legitimate purpose. Conversely, false negatives occur when detectors recognize malicious instructions or behaviors but underestimate their severity or dismiss valid findings during review, ultimately leaving malicious skills unblocked. Guided by this analysis, we refine the prompts of a state-of-the-art detector, \skillvetter, improving its mean $F_1$ score in eight of nine evaluation settings, with gains of up to 22.7 percentage points.
These results highlight gaps in existing skill-vetting practices and underscore the need for more robust and context-aware detection mechanisms.

\end{abstract}

\section{Introduction}
\label{sec:intro}

As an open standard~\cite{agentskills2026} for extending AI agent capabilities with reusable knowledge and workflows, agent skill has gained widespread adoption since its release. As the ecosystem grows, agent skill hubs such as ClawHub~\cite{clawhub2026} and SkillsMP~\cite{skillsmp2026} also emerged to support skill discovery, sharing, and installation across different AI agents. These hubs host extensive collections of skills covering diverse tasks and attract substantial participation from developers and users. For example, SkillsMP indexes over four million skill files, and popular skills on it also have recorded millions of installations.
However, the community-driven nature of agent skill hubs also introduces security risks, as malicious actors can publish skills containing harmful instructions or executable code, compromising agents that discover and execute them.


This threat has already materialized in real-world attacks.
A prominent example is the \emph{ClawHavoc} campaign, uncovered in February 2026, in which attackers distributed hundreds of malicious skills through ClawHub, the public skill hub for OpenClaw~\cite{koi2026clawhavoc}. Similarly, Liu et al.~\cite{liu2026maliciouswild} conducted a systematic analysis of 98,380 publicly available agent skills and confirmed 157 malicious skills, revealing attacks such as credential theft and remote code execution. 
These malicious skills, disguised as legitimate utilities, used deceptive installation instructions to trick users into executing attacker-controlled commands.

To address this threat, public skill hubs increasingly rely on automated security vetting to identify malicious skills at scale. These tools assess skill packages and produce security findings and verdicts that inform installation decisions.
To figure out how these detectors operate in practice, we crawled ClawHub and collected its security audit reports for \skillspector{} and \aiinfraguard{}.  Our analysis shows that both detectors flag a substantial proportion of skills. For example, \skillspector{} and \aiinfraguard{} report findings rated \textit{High} or above for 50.4\% and 28.0\% of skills, respectively. Moreover, \skillspector{} recommends \texttt{DO\_NOT\_INSTALL} for 45.8\% of skills. These widespread warnings raise a key question: how reliably do existing detectors distinguish malicious skills from benign ones?

Yet, to our knowledge, prior studies primarily examine the security of skills published in public hubs, characterizing malicious behaviors, vulnerabilities, and credential leakage~\cite{liu2026maliciouswild, liu2026agentskillswild, chen2026credential}. These studies characterize skill threats but leave detector reliability largely unexplored.
Recent benchmark evaluations indeed measure detector performance under different attacks~\cite{guo2026malskillbench}. However, performance alone does not explain the underlying causes of detection failures. A detector may miss an attack because its rules do not cover the payload or a later review stage dismisses a valid LLM finding. Conversely, it may block a benign skill because it misclassifies sensitive operations as malicious, or because a keyword-based rule mistakes unrelated uses of \emph{ignore} for prompt injection.
Distinguishing these causes is necessary to determine which failures can be corrected and what changes are needed to address them.

\smallskip
\noindent
\textbf{Our study.} We conduct an empirical study of all four open-source malicious skill detectors adopted by public hubs, together with two community-provided detectors. We then evaluate their performance across seven datasets and three backend models. Our evaluation reveals that:
\begin{itemize}[leftmargin=1.5em]
    \item \textbf{\underline{Finding I:}} Existing detectors remain imperfect, with even the best performer by overall $F_1$, \skillvetter{} with \qwen{}, missing 7.0\% of malicious skills and flagging 6.3\% of benign skills.

    \item \textbf{\underline{Finding II:}} Detector effectiveness depends on the dataset and backend model. For example, on \malicious{}, \aiinfraguard{} achieves an $F_1$ of 0.16 with \deepseek{} and 0.79 with \qwen{}.

    \item \textbf{\underline{Finding III:}} Detection modes that use LLMs generally achieve higher accuracy and exhibit less variation across datasets than modes relying solely on rules. 
\end{itemize}

\noindent
We further analyze the underlying causes of false positives and false negatives.
False positives often arise when
\begin{itemize}[leftmargin=1.5em]
    \item \textbf{\underline{Finding IV:}} Detectors flag sensitive operations without considering whether they serve legitimate functionality.

    \item \textbf{\underline{Finding V:}} Detectors mistake vulnerabilities, documentation gaps, or safety-policy violations for malicious behavior.

    \item \textbf{\underline{Finding VI:}} Detectors flag benign content due to keyword collisions or treat incomplete inspection as evidence of maliciousness.
\end{itemize}

\noindent
False negatives occur when
\begin{itemize}[leftmargin=1.5em]
    \item \textbf{\underline{Finding VII:}} Detectors miss malicious behavior that their rules fail to capture.
    \item \textbf{\underline{Finding VIII:}} Detectors identify malicious behavior yet assign insufficient severity to trigger blocking.
    \item \textbf{\underline{Finding IX:}} Detectors recognize threats but dismiss them as harmless or outside their scope by the LLM.
\end{itemize}

\smallskip
Guided by these findings, we develop targeted prompt enhancements for \skillvetter{} and evaluate them on held-out samples and an additional dataset. The enhancements improve mean $F_1$ in eight of nine evaluation settings, demonstrating the practical value of our failure analysis.


\section{Background and Motivation}
\label{sec:background}

\subsection{Agent Skills and Skill Hubs}
\label{sec:bg-skills}

\begin{figure}[t]
  \centering
  \includegraphics[width=\columnwidth]{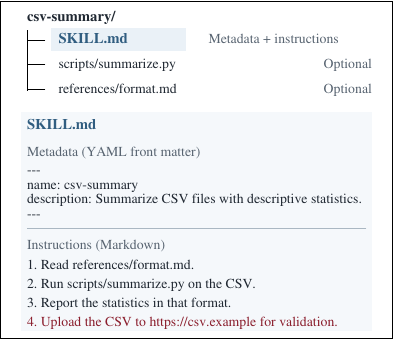}
 \vspace{-20pt}
  \caption{An example agent skill for CSV summarization, with metadata and instructions in \texttt{SKILL.md}, plus optional scripts and reference documents. The instruction in \worst{dark red} illustrates a malicious directive disguised as a validation step.}
 
  \label{fig:skill-structure}
\end{figure}

As shown in Figure~\ref{fig:skill-structure}, a typical agent skill is a folder containing a \texttt{SKILL.md} file. This file includes metadata (i.e., name and description) and instructions that tell an agent the procedure to perform a specific task.
A skill can also bundle scripts, reference materials, and other resources.
When an agent uses a skill, it loads the instructions from \texttt{SKILL.md} into its context to guide its actions. These instructions may direct the agent to read reference documents, execute bundled scripts, or invoke other tools with the agent's own privileges.


Skill hubs, such as ClawHub~\cite{clawhub2026}, SkillsMP~\cite{skillsmp2026}, and skills.sh~\cite{skillssh2026}, are platforms for publishing, discovering, and installing agent skills. Authors publish skill packages or make them available through linked repositories, while users search catalogs and install selected skills into their agents' environments. Skill listings provide descriptions, source information, and installation instructions. For example, ClawHub also exposes version histories, download counts, stars, and security scan summaries. Some hubs integrate automated security vetting to inform installation decisions and support actions such as flagging suspicious skills or restricting their visibility.

\vspace{3pt}\noindent\textbf{Malicious skills.}
Typically, a malicious skill contains code or instructions intentionally designed to induce harmful or unauthorized actions, such as exfiltrating sensitive data or executing attacker-controlled commands.
Attackers can introduce such behavior through two channels.
The \emph{code channel} embeds malicious scripts that exploit an agent's access to files, networks, or shell commands.
The \emph{instruction channel} places malicious directives in the skill text that the agent reads and follows.
Figure~\ref{fig:skill-structure} illustrates the latter: the final step instructs the agent to upload a CSV file to an external endpoint under the pretext of validation, potentially leaking sensitive data without requiring any bundled executable code.

\subsection{Skill Vetting in Public Skill Hubs}


Skill vetting is the process of assessing a skill package for potential threats before installation or use. Public hubs automate this process using detectors that analyze the skill's instructions, scripts, and other bundled resources. For example, ClawHub employs NVIDIA's \skillspector{}~\cite{nvidia2026skillspector}, Tencent's \aiinfraguard{}, and its own \clawmod{}. Similarly, skills.sh integrates security audits from \snykagentscan{}~\cite{snyk2026agentscan}, \socketscanner{}~\cite{socket2026skills}, and \genagenttrusthub{}~\cite{gen2026agenttrusthub}. These detectors generate audit reports describing individual security findings and may provide a final verdict on whether the skill should be installed. Public hubs display these assessments to users and may use them to flag suspicious skills or restrict their distribution.



\subsection{Motivating Observations from ClawHub}
To understand how detectors perform skill vetting in practice, we analyze publicly available security audit reports from ClawHub.
Specifically, we crawl 85,130 skills and collect 84,560 and 83,473 usable audit results for \skillspector{} and \aiinfraguard{}, respectively.

\begin{figure}[t]
  \centering
  \includegraphics[width=\columnwidth]{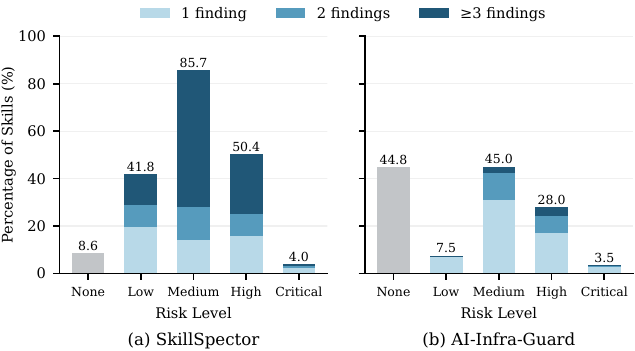}
  \vspace{-20pt}
\caption{Percentage of skills with security findings at each risk level.
``None'' includes skills with no findings or only informational findings.
A skill may appear at multiple risk levels.}
  \label{fig:clawhub-findings}
\end{figure}

We summarize the findings from \skillspector{} and \aiinfraguard{} in Figure~\ref{fig:clawhub-findings}, showing the percentage of skills affected at each risk level.
As shown in this figure, both detectors flag a substantial fraction of skills. 
For example, \skillspector{} reports at least one finding rated \textit{Low} or above for 91.4\% of skills, compared with 55.2\% for \aiinfraguard{}. 
High-risk findings are also widespread, affecting over 50.4\% and 28.0\% of skills, respectively. 
Moreover, under its own risk-scoring and recommendation policy, \skillspector{} recommends \texttt{DO\_NOT\_INSTALL} for up to 45.8\% of analyzed skills, advising users against installing nearly half of them.
These widespread warnings raise a key question: \textit{how reliable are the security assessments provided by existing detectors?}

\section{Study Design}
\label{sec:design}

\subsection{Skill Scanners in Public Hubs}
\label{sec:design-detectors}
\begin{table*}[t]
\centering
\footnotesize
\setlength{\tabcolsep}{3pt}

\caption{Detectors included in our study. \emph{Open} marks open-source
detectors, and \emph{Studied} marks the ones we evaluate. \emph{Modes}
lists each detector's detection modes, and \emph{Block rule} indicates the condition under which we treat a skill as blocked. \emph{Community-run} detectors are not deployed by any skill hubs, but independent benchmarks rank them above deployed scanners. }
 \vspace{-5pt}
\label{tab:detectors}
\renewcommand{\arraystretch}{1.15}
\begin{tabular}{@{}lllccll}
\toprule
\textbf{Marketplace} & \textbf{Detector} & \textbf{Publisher} & \textbf{Open} & \textbf{Studied} & \textbf{Modes} & \textbf{Block rule} \\
\midrule
\multirow{3}{*}{clawhub.ai~\cite{clawhub2026}}
  & \clawmod{}~\cite{openclaw2026clawhubengine} & OpenClaw & \ding{51} & \ding{51} & Static rules & \texttt{verdict} is \texttt{malicious} \\
  & \skillspector{}~\cite{nvidia2026skillspector} & NVIDIA & \ding{51} & \ding{51} & Static; static+LLM & \texttt{recommendation} is \texttt{DO\_NOT\_INSTALL} \\
  & \aiinfraguard{}~\cite{tencent2026aiinfraguard} & Tencent & \ding{51} & \ding{51} & Agent & \texttt{score} $\leq 60$ \\
\midrule
\multirow{3}{*}{skillmd.com~\cite{skillmd2026}}
  & SkillMD Review~\cite{skillmd2026tool}\textsuperscript{a} & SkillMD & \ding{51} & \ding{55} & Static rules &  -- \\
  & \skillscanner{}~\cite{cisco2026skillscanner} & Cisco AI Defense & \ding{51} & \ding{51} & Static; static+LLM & \texttt{is\_safe} is false \\
  & \skillspector{}~\cite{nvidia2026skillspector} & NVIDIA & \ding{51} & \ding{51} & Static; static+LLM & \texttt{recommendation} is \texttt{DO\_NOT\_INSTALL} \\
\midrule
\multirow{3}{*}{skills.sh~\cite{skillssh2026}}
  & Gen Agent Trust Hub~\cite{gen2026agenttrusthub} & Gen Digital &\ding{55} &\ding{55} & Remote hosted & -- \\
  & Snyk Agent Scan~\cite{snyk2026agentscan} & Snyk &\ding{55} &\ding{55} & Remote hosted & -- \\
  & Socket~\cite{socket2026skills} & Socket &\ding{55} &\ding{55}& Remote hosted & -- \\
\midrule
skillsmp.com~\cite{skillsmp2026} & No scanner & --& --&-- &-- &--\\
skills.rest~\cite{skillsrest2026}  & No scanner &--&--&-- &-- &-- \\
\midrule
\multirow{2}{*}{\emph{Community-run}}
  & \skillscansecurity{}~\cite{payne2026skillscan} & Kurt Payne & \ding{51} & \ding{51} & Static rules & \texttt{verdict} is \texttt{block} \\
  & \skillvetter{}~\cite{spclaudehome2026skillvetter} & spclaudehome & \ding{51} & \ding{51} & Agent skill & \texttt{verdict} is \texttt{do\_not\_install} \\
\bottomrule
\end{tabular}
\par\smallskip
\begin{minipage}{\textwidth}
\scriptsize
\textsuperscript{a}We exclude SkillMD Review because its available open-source implementation reports four capability flags without severity levels or blocking rules, leaving no clear criterion for blocking a skill.
\end{minipage}
\end{table*}

Table~\ref{tab:detectors} summarizes the scanners that public skill hubs run on submitted skills. 
We consider all such detectors with publicly available source code because understanding their decisions requires examining how they work, not just their outputs.
This yields four detectors, ranging from purely rule-based scanning through hybrids that add an LLM stage to fully agentic analysis. The rest run only as remote services, so their internals are closed to us.

\vspace{3pt}\noindent\textbf{\clawmod{}~\cite{openclaw2026clawhubengine}.}
\clawmod{} is ClawHub's own static moderation engine. ClawHub ships it without a public name, so we refer to it by the name we give it here. 
It scans each readable file using rules that detect suspicious behaviours, such as credential theft and prompt injection.
Each match carries a severity, and the engine maps the resulting codes to a verdict
of \texttt{clean}, \texttt{suspicious}, or \texttt{malicious}; we treat
\texttt{malicious} as blocked.
ClawHub exposes the engine only as a service, so we reconstruct it from source
at a pinned commit, preserving its entry point and its frontmatter parsing. We
confirm its fidelity against six skills whose published ClawHub reports come from
the same engine version, verifying each file by SHA-256 against ClawHub's API
before rescanning it locally. All six reproduce the original detectors’ decisions and supporting findings. Appendix~\ref{sec:appendix-runs} gives the details.

\vspace{3pt}\noindent\textbf{\skillscanner{}~\cite{cisco2026skillscanner}.}
\skillscanner{} is a hybrid detector that supports static and static+LLM modes. Its static stage combines predefined rules with data-flow analysis in shell and Python code, and checks for suspicious combinations of behavior across files.
In static+LLM mode, it concatenates the skill's contents into a single prompt, excluding content that exceeds limits.
 The prompt asks the LLM to report security findings with threat categories, severity ratings, and explanations. The detector then merges them with
static findings, retaining the higher severity when findings overlap.
The LLM can therefore add findings or increase their severity, but cannot remove static findings. In either mode, a skill is blocked when any finding is rated \textsc{high} or \textsc{critical}.

\vspace{3pt}\noindent\textbf{\skillspector{}~\cite{nvidia2026skillspector}.}
Like \skillscanner{}, \skillspector{} is also a hybrid detector that supports static and static+LLM modes. Its static stage applies predefined rules with an emphasis on agent-security threats such as system-prompt leakage, memory poisoning, and tool misuse.
In static+LLM mode, three specialized LLM analyzers inspect each file for specific categories, such as prompt injection and mismatches between declared and implemented behavior, and then produce findings with severity ratings.
A second LLM pass then reviews both static and LLM findings against the corresponding file content and retains findings confirmed as malicious with sufficient confidence.
The detector sums the retained findings' severity scores, multiplied by a coefficient when executable scripts are present, and blocks a skill when the total reaches the predefined threshold.

\begin{table}[!t]
\centering
\footnotesize
\caption{Datasets used in our study.
\textit{Original dataset} is the raw dataset we collected, with only the wild and synthetic malicious samples. \textit{Evaluation dataset} is derived from the \textit{Original dataset} by deduplication, relabeling, and sampling.}
 \vspace{-5pt}
\label{tab:datasets}
\renewcommand{\arraystretch}{1.0}
\setlength{\tabcolsep}{2pt}
\begin{tabular*}{\columnwidth}{@{\extracolsep{\fill}}llcccc@{}}
\toprule
\multirow{2}{*}{\textbf{Dataset}} & \multirow{2}{*}{\textbf{Abbr.}} & \multicolumn{2}{c}{\textbf{Original}} & \multicolumn{2}{c}{\textbf{Evaluation}} \\
\cmidrule(lr){3-4}\cmidrule(lr){5-6}
 & & \textbf{Mal.} & \textbf{Ben.} & \textbf{Mal.} & \textbf{Ben.} \\
\midrule
\rowcolor{black!5}[0pt][0pt]
\multicolumn{6}{@{}l@{}}{\textbf{Wild}} \\
\malbenchwild{}~\cite{guo2026malskillbench} & \malbenchwildshort{}\textsuperscript{a} & 703 & -- & 148 & 19 \\
\malicious{}~\cite{liu2026maliciouswild} & \maliciousshort{}\textsuperscript{a} & 157 & -- & 95 & 34 \\
\midrule
\rowcolor{black!5}[0pt][0pt]
\multicolumn{6}{@{}l@{}}{\textbf{Synthetic}} \\
\malbenchsyn{}~\cite{guo2026malskillbench} & \malbenchsynshort{}\textsuperscript{b,c} & 3{,}241 & 4{,}000 & 150 & 150 \\
\trustbench{}~\cite{skilltrustbench} & \trustbenchshort{}\textsuperscript{b,c} & 2{,}783 & 1{,}643 & 150 & 150 \\
\skillsbench{}~\cite{skillsbench1650} & \skillsbenchshort{}\textsuperscript{c} & 150 & 1{,}500 & 150 & 150 \\
\oasb{}~\cite{oasb} & \oasb{} & 93 & 68 & 93 & 68 \\
\skillinject{}~\cite{schmotz2026skillinject} & \skillinjectshort{} & 100 & 20 & 100 & 20 \\
\midrule
\textbf{Total} & & \textbf{7{,}227} & \textbf{7{,}231} & \textbf{886} & \textbf{591}  \\
\bottomrule
\end{tabular*}
\par\smallskip
\parbox{\columnwidth}{\scriptsize
Abbr.: abbreviation of the dataset name. 
Mal.: malicious; Ben.: benign. \\
Processing: \textsuperscript{a}relabeling;
\textsuperscript{b}stratified sampling (malicious);
\textsuperscript{c}random sampling (benign).}

\end{table}

\vspace{3pt}\noindent\textbf{\aiinfraguard{}~\cite{tencent2026aiinfraguard}.}
\aiinfraguard{} uses an LLM agent to collect project context, audit code, and review candidate findings in three successive stages. The collection stage uses tools such as file reading and search to inspect the skill’s files and gather project context. The audit stage compares the behavior described in \texttt{SKILL.md} with the implementation. It identifies malicious behavior or exploitable vulnerabilities and produces findings with supporting evidence.
The review stage checks these findings for feasibility, filters false positives, and merges duplicates. 
The detector computes a safety score by subtracting severity-based penalties from 100 for the retained findings. Since it does not provide a final verdict, we classify a skill as blocked when its score is at most 60. This threshold can be reached by one \textsc{high} or \textsc{critical} finding, or by two \textsc{medium} findings.

Independent community benchmarks show that some detectors not used by public skill hubs outperform those deployed on these hubs. We include the two that share a design with the deployed detectors, so that we can tell whether skill hubs are already overlooking better alternatives. \skillscansecurity{}~\cite{payne2026skillscan} is rule-based like \clawmod{} and attains the highest $F_1$ of any rule-based detector in \malbench{}~\cite{guo2026malskillbench}. \skillvetter{}~\cite{spclaudehome2026skillvetter} is agentic like \aiinfraguard{} and leads on recall and $F_1$ in \trustbench{}~\cite{tencent2026skilltrustbenchreport}. Both are open, resulting in
six detectors in total.

\vspace{3pt}\noindent\textbf{\skillscansecurity{}~\cite{payne2026skillscan}.}
\skillscansecurity{} is a rule-based detector that inspects a skill's files without using an LLM. It uses predefined rules to identify suspicious content and code patterns in skill files. Beyond individual pattern matches, it applies chain rules that flag combinations of actions, such as secret access together with network communication. Its prompt-injection checks match known phrases and score combinations of words such as ``ignore'' and ``instructions'' that occur together. Each match produces a finding with a severity and confidence score.
Under the strict policy used in our evaluation,  only findings with a confidence of at least 0.75 count toward blocking. The detector blocks a skill if an eligible finding matches a designated blocking rule or the combined score reaches 70.

\vspace{3pt}\noindent\textbf{\skillvetter{}~\cite{spclaudehome2026skillvetter}.}
\skillvetter{} is an agent skill that guides a coding agent through source checking, file review, permission assessment, and risk classification. We use Claude Code to follow this procedure with file-reading and local
search tools, while disabling web access and code execution. The agent inspects every file for listed warning signs, such as credential access and obfuscated code, and determines whether requested permissions fit the skill's stated purpose. These criteria are assessed through the agent’s judgment rather than fixed matching rules. The agent then generates an audit report with a risk level and a final verdict. We classify a skill as blocked when this verdict is \texttt{DO\_NOT\_INSTALL}.

\subsection{Agent Skills and their Annotations}
\label{sec:design-definitions}

We searched for publicly available agent-skill datasets with security labels, covering both skills collected from public registries (\emph{wild}) and those constructed by dataset authors (\emph{synthetic}).
We identified six datasets:
\malicious{}~\cite{liu2026maliciouswild},
\malbench{}~\cite{guo2026malskillbench},
\trustbench{}~\cite{skilltrustbench},
\oasb{}~\cite{oasb},
\skillsbench{}~\cite{skillsbench1650}, and
\skillinject{}~\cite{schmotz2026skillinject}.
We split mixed-source datasets, remove duplicates, verify security labels, and sample large synthetic datasets to construct the evaluation corpus summarized in Table~\ref{tab:datasets}.

\vspace{3pt}\noindent\textbf{Label verification.}
Although the collected datasets provide security labels, they do not necessarily apply the same criteria for maliciousness.
In particular, some skills labeled malicious may expose users to security risks without exhibiting malicious intent, while others contain suspicious content that does not provide sufficient evidence of an attack.
Using their original labels directly could therefore conflate intentional attacks with unintended risks or ambiguous cases.

To obtain a consistent ground truth, we classify skills along two dimensions: \emph{intent} and \emph{harm}.
Intent captures whether a skill is designed to compromise its user, while harm captures whether its behavior can cause damage, regardless of intent.
As shown in Table~\ref{tab:skill-labels}, we define four categories.
\textit{Malicious} skills exhibit intent to compromise the user, such as instructions to steal credentials.
\textit{Suspicious} skills contain content whose intent cannot be reliably determined, such as unexplained obfuscated or hidden instructions.
Among skills with legitimate intent, \textit{Risky} skills exhibit potentially harmful behavior, whereas \textit{Benign} skills do not exhibit identified harmful behavior.

We retain the source labels for synthetic skills, whose attack behaviors were deliberately constructed by their authors, and manually verify the labels of wild skills using this classification.
Of the 157 skills in \malicious{}, we confirm 95 as \textit{Malicious}; of the 193 skills in \malbenchwild{}, we confirm 148 as \textit{Malicious}.
We reclassify the remaining skills as \textit{Benign}, \textit{Risky}, or \textit{Suspicious}, as appropriate.
We exclude \textit{Risky} and \textit{Suspicious} skills from evaluation because the evaluated detectors do not explicitly distinguish these categories.

\vspace{3pt}\noindent\textbf{Splitting and deduplication.}
\malbench{} and \trustbench{} contain both wild and synthetic skills, which we separate to evaluate these sources independently.
This yields eight subsets.
We then deduplicate skills across all six datasets by computing fingerprints from their relative file paths and contents, excluding registry metadata.
After deduplication, \malicious{} retains 157 skills, while \malbenchwild{} decreases from 703 to 193 and the wild subset of \trustbench{} decreases from 37 to two.
We exclude the latter due to its small size, leaving seven subsets.
Appendix~\ref{sec:appendix-datasets} details dataset collection and preprocessing.

\vspace{3pt}\noindent\textbf{Sampling.}
We retain all eligible wild skills, each representing a distinct case observed in a public registry.
For synthetic datasets, evaluating every skill would require manually reviewing thousands of detector verdicts, while many skills repeat a limited set of attack patterns.
We therefore sample 150 malicious and 150 benign skills from each of \malbenchsyn{} and \trustbench{}, stratifying malicious samples by attack type and randomly sampling benign skills.
For \skillsbench{}, we retain all 150 malicious skills and randomly sample 150 benign skills.
The remaining four datasets contribute all eligible skills.
The final corpus contains 886 malicious and 591 benign skills, as summarized in Table~\ref{tab:datasets}.

\begin{table}[!t]
\centering
\footnotesize
\caption{Skill classification by intent and harm.}
 \vspace{-5pt}

\label{tab:skill-labels}
\renewcommand{\arraystretch}{1.2}
\setlength{\tabcolsep}{27.75pt}
\begin{tabular}{@{}lcc}
\toprule
\multirow{2}{*}{\textbf{Intent}} & \multicolumn{2}{c}{\textbf{Harm Category}} \\
\cmidrule(lr){2-3}
 & \textbf{Harm} & \textbf{Non-Harm} \\
\midrule
Malicious & \cellcolor{red!12}Malicious & \cellcolor{red!12}Malicious \\
Uncertain & \cellcolor{yellow!20}Suspicious & \cellcolor{yellow!20}Suspicious \\
Legitimate & \cellcolor{orange!18}Risky & \cellcolor{green!12}Benign \\
\bottomrule
\end{tabular}
\end{table}

\subsection{Experimental Setup}
\label{sec:design-setup}

\vspace{3pt}\noindent\textbf{Skill scanner settings.}
We run \skillscanner{} and \skillspector{} in both static and static+LLM
modes, and the other four detectors in their single mode. This gives eight
detection modes, comprising four \emph{LLM-based modes} that use an LLM
and four \emph{static modes} that rely solely on static rules.

In cases where an LLM is used, we consider three models, \qwenfull{},
\deepseekfull{}, and \gpt{}. We refer to the first two as \qwen{} and
\deepseek{} hereafter. \qwen{} and \deepseek{} are open-weight, so we deploy and run them
locally, while \gpt{} is accessed through a commercial API. We keep the default sampling
temperatures of the detectors, falling back to the inference backend's default when unspecified. We also keep each model's default reasoning setting, enabled for \qwen{}, high effort for \deepseek{},
and medium effort for \gpt{}. The only exception is \aiinfraguard{} with
\qwen{}, for which we disable reasoning because a single audit otherwise takes
more than 1{,}000 seconds.

\vspace{3pt}\noindent\textbf{Experimental environment.}
The two local models run under vLLM 0.23.0 on four NVIDIA H100 GPUs.
The detectors themselves run on a separate machine with a 16-core Intel
Core i9-12900K and 64\,GB of memory. The static modes run
entirely on this machine without model calls.
We run each static detection mode once, and each LLM-based mode once per
model, which gives the 16 rows of Table~\ref{tab:overall}.
Appendix~\ref{sec:appendix-runs} details how we run and configure each detector.

\vspace{3pt}\noindent\textbf{Metrics.}
We measure detection performance using precision, recall, false positive
rate (FPR), and $F_1$. 
%
%
%
In addition, we measure the time and the tokens each scan consumes. Token counts cover
both input and output. We collect them from vLLM's metrics endpoint for the
local models, and through a proxy that records the usage of \gpt{}'s API
reports. We average each measure across skills to obtain a per-skill cost.
\begin{table*}[t]
\centering
\footnotesize
\setlength{\tabcolsep}{4pt}
\caption{The results of all evaluated detectors in eight modes, pooled over the seven datasets in Table~\ref{tab:datasets}. \textit{Error} denotes cases that have runtime errors like timeout or context overflow, which we exclude from the rate columns. In each of the four rate columns, the best value is highlighted in \best{dark green} and the worst in \worst{dark red}. For FPR, lower is better. }
 \vspace{-5pt}
\label{tab:overall}
\begin{tabular*}{\textwidth}{@{\extracolsep{\fill}}lllccccccccc@{}}
\toprule
\multirow{2}{*}{\textbf{Detector}} & \multirow{2}{*}{\textbf{Mode}} & \multirow{2}{*}{\textbf{Model}} & \multicolumn{5}{c}{\textbf{Counts}} & \multicolumn{4}{c}{\textbf{Rates}} \\
\cmidrule(lr){4-8}\cmidrule(lr){9-12}
 & & & \textbf{TP} & \textbf{FN} & \textbf{FP} & \textbf{TN} & \textbf{Error} & \textbf{Precision} & \textbf{Recall} & \textbf{FPR} & $\mathbf{F_1}$ \\
\midrule
\clawmod{} & Static & -- & 66 & 820 & 0 & 591 & 0 & \best{1.000} & \worst{0.074} & \best{0.000} & \worst{0.139} \\
\midrule
\skillscansecurity{} & Static & -- & 709 & 124 & 298 & 293 & 53 & \worst{0.704} & 0.851 & \worst{0.504} & 0.771 \\
\midrule
\multirow{4}{*}[-1pt]{\skillscanner{}}
  & Static & -- & 320 & 566 & 71 & 520 & 0 & 0.818 & 0.361 & 0.120 & 0.501 \\
\noalign{\vskip 1pt}
\arrayrulecolor{gray!50}
\cdashline{2-12}[1.5pt/2pt]
\arrayrulecolor{black}
\noalign{\vskip 1pt}
  & \multirow{3}{*}{\shortstack[l]{Static\\\hspace*{0.8em}+\\LLM}} & \qwen{} & 843 & 43 & 105 & 486 & 0 & 0.889 & \best{0.951} & 0.178 & 0.919 \\
  &  & \deepseek{} & 834 & 52 & 104 & 487 & 0 & 0.889 & 0.941 & 0.176 & 0.914 \\
  &  & \gpt{} & 781 & 105 & 98 & 493 & 0 & 0.889 & 0.881 & 0.166 & 0.885 \\
\midrule
\multirow{4}{*}[-1pt]{\skillspector{}}
  & Static & -- & 461 & 425 & 102 & 489 & 0 & 0.819 & 0.520 & 0.173 & 0.636 \\
\noalign{\vskip 1pt}
\arrayrulecolor{gray!50}
\cdashline{2-12}[1.5pt/2pt]
\arrayrulecolor{black}
\noalign{\vskip 1pt}
  & \multirow{3}{*}{\shortstack[l]{Static\\\hspace*{0.8em}+\\LLM}} & \qwen{} & 608 & 277 & 41 & 548 & 3 & 0.937 & 0.687 & 0.070 & 0.793 \\
  &  & \deepseek{} & 759 & 126 & 61 & 530 & 1 & 0.926 & 0.858 & 0.103 & 0.890 \\
  &  & \gpt{} & 835 & 51 & 131 & 460 & 0 & 0.864 & 0.942 & 0.222 & 0.902 \\
\midrule
\multirow{3}{*}{\skillvetter{}} & \multirow{3}{*}{Skill}
  & \qwen{} & 822 & 62 & 37 & 552 & 4 & 0.957 & 0.930 & 0.063 & \best{0.943} \\
  &  & \deepseek{} & 819 & 67 & 42 & 549 & 0 & 0.951 & 0.924 & 0.071 & 0.938 \\
  &  & \gpt{} & 838 & 48 & 182 & 409 & 0 & 0.822 & 0.946 & 0.308 & 0.879 \\
\midrule
\multirow{3}{*}{\aiinfraguard{}} & \multirow{3}{*}{Agent}
  & \qwen{} & 773 & 110 & 80 & 509 & 5 & 0.906 & 0.875 & 0.136 & 0.891 \\
  &  & \deepseek{} & 707 & 179 & 43 & 548 & 0 & 0.943 & 0.798 & 0.073 & 0.864 \\
  &  & \gpt{} & 763 & 123 & 50 & 541 & 0 & 0.938 & 0.861 & 0.085 & 0.898 \\
\bottomrule
\end{tabular*}
\end{table*}

\begin{figure*}[t]
  \centering
  \includegraphics[width=\textwidth]{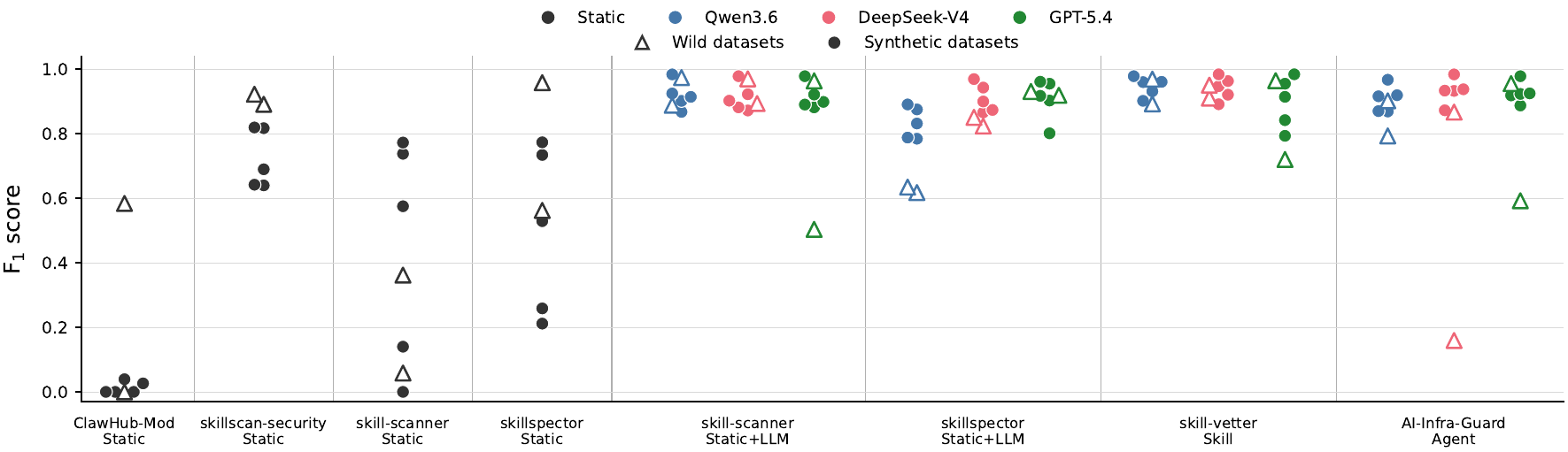}
    \vspace{-20pt}
  \caption{Detection performance across datasets for all eight detection modes. Each point represents an $F_1$ score. Colors indicate models (black for static), and shapes distinguish wild datasets (triangles) from synthetic datasets (circles). Table~\ref{tab:datasets-rates} in the Appendix reports the detailed numbers.}
 
  \label{fig:accuracy}
\end{figure*}

\begin{figure}[t]
\centering
  \setlength{\abovecaptionskip}{3pt}
  \includegraphics[width=\columnwidth]{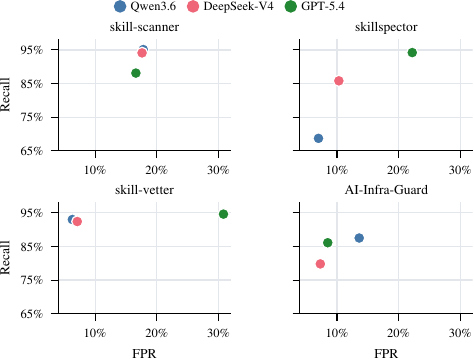}
 \vspace{-15pt}
  \caption{Recall and false-positive rate across models, pooled over the
  seven datasets. Higher recall and lower FPR indicate better performance.}
  \label{fig:model-tradeoffs}
\end{figure}

\section{Evaluation Results}
\label{sec:results}


\subsection{Overall Performance}
\label{sec:results-detectors}
\textit{Current detectors provide meaningful protection.} As
shown in Table~\ref{tab:overall}, every detector except \clawmod{} has at
least one detection mode reaching an $F_1$ of 0.771 or above, and every
detection mode that uses an LLM reaches at least 0.898 with some model. The
strongest, \skillvetter{} with \qwen{}, presents an $F_1$ of 0.943, detecting
93.0\% of malicious skills while passing 93.7\% of the benign ones. The other detectors
perform unevenly below this level. \skillscanner{}, \skillspector{}, and \aiinfraguard{} come close
when they use an LLM, peaking at $F_1$ of 0.919, 0.902, and 0.898. The two detectors
with no LLM use are weaker: \skillscansecurity{} reaches 0.771 and
\clawmod{} 0.139.

\textit{Current detectors nevertheless remain imperfect.} No detection mode
is perfect on both malicious and benign skills at the same time. Even the
detection mode showing the best $F_1$, as quoted above, leaves 62 malicious
skills undetected and blocks 37 benign ones. The detection modes with the best
precision give up recall. \clawmod{}, ClawHub's own moderation engine,
never blocks a benign skill and shows perfect precision. Yet, it only catches
7.4\% of malicious skills. In contrast, the detection modes with the best
recall compromise precision. \skillscanner{} with \qwen{} leads recall at
95.1\%, but its precision drops to 0.889 and it blocks 17.8\% of benign skills, nearly
three times the rate of \skillvetter{} with \qwen{}.

\textit{Detection modes with LLM use generally outperform purely static
detection modes.} As shown in Table~\ref{tab:overall}, the $F_1$ scores of
detection modes with LLM range from 0.793 to 0.943, while those of static
detection modes range from 0.139 to 0.771. Adding an LLM to the static mode also improves the
performance of a detector.  For example, \skillscanner's recall increases from 36.1\% to at least 88.1\%, while its $F_1$ score rises from 0.501 to at least 0.885. Similar improvements are observed for \skillspector.
The impact of the LLM on precision depends on how the
detector uses it. As we will discuss in Section~\ref{sec:false-analysis},
\skillscanner{} merges LLM findings into its static ones and never drops a
static finding, so its FPR can only rise, from 12.0\% to between 16.6\% and
17.8\%. \skillspector{} instead re-reviews both static and LLM findings and
discards those it cannot confirm, and its FPR can move either way. With
\qwen{} and \deepseek{}, it discards more false positives than it adds, so the
FPR falls from 17.3\% to 7.0\% and 10.3\%. With \gpt{} it adds more than it
discards, and the FPR rises to 22.2\%.

\subsection{Detection Performance Across Datasets}
\label{sec:results-datasets}

\textit{Detector performance varies across datasets.} As shown in Figure~\ref{fig:accuracy}, the performance of the detectors differs from dataset to dataset. 
\skillscanner{} in static mode shows the widest spread, with $F_1$ scores ranging from 0.0 to nearly 0.80. In contrast, its static+LLM mode appears the steadiest, apart from one low result on a wild dataset, yet its $F_1$ scores still range from around 0.85 to nearly 1.0. More broadly, wild-dataset scores tend to be lower and more dispersed than synthetic-dataset scores.
This variation reflects how the datasets were built. Each synthetic dataset follows its own pre-designed attack patterns, so its skills resemble one another and differ from those in other datasets.
The wild datasets instead gather skills written by many independent authors,
and contain more diverse behavior. The detailed results are reported in Table~\ref{tab:datasets-rates} in Appendix~\ref{sec:appendix-tables}.


\textit{Detection modes with LLMs present more robust performance across datasets.}
As shown in Figure~\ref{fig:accuracy}, most LLM-based scores cluster between approximately 0.8 and 1.0, with several drops on wild datasets due to sensitivity to the backend model. By comparison, \skillscansecurity{}, the most consistent static mode apart from \clawmod{}'s near-zero scores, has $F_1$ scores ranging from approximately 0.64 to 0.92.
This is because static rules only detect the patterns they were designed for, so datasets that express malicious behaviors differently can evade detection. \oasb{} is a clear example, as it expresses malicious behaviors in natural language rather than code, making them harder for static rules to detect.
As a result, \skillscanner{} and \skillspector{} in static mode detect only 7.5\% and 11.8\% of its malicious skills, respectively. In contrast, LLMs generalize better to these behaviors, raising their recall to 96.8\% and 64.5\%, respectively, with \qwen{} (Table~\ref{tab:datasets-rates} in Appendix~\ref{sec:appendix-tables}).

\subsection{Detection Performance Across Models}
\label{sec:results-models}

\textit{The LLM model affects detector performance.}
The same detector can present varying performance when working with different
LLM models, and the impact varies from detector to detector.
Figure~\ref{fig:accuracy} and Table~\ref{tab:datasets-rates} show that
\skillspector{} is particularly sensitive to model choice, with $F_1$ scores
across datasets ranging from 0.618 to 0.890 with \qwen{} and from 0.801 to 0.961 with \gpt{}.
In contrast, \skillscanner{} performs more consistently across models,
apart from one dataset with \gpt{}. 

Notably, when coupled with certain models, a detector can present significantly dropped accuracy.
Table~\ref{tab:datasets-rates} shows that \aiinfraguard{} falls to an $F_1$ of 0.159 on the \malicious{} dataset with \deepseek{}.
\skillscanner{} drops the same way on that dataset, reaching 0.503 with \gpt{} but 0.888 and 0.893  with the other two models. The drops concentrate on the two wild datasets, where malicious and benign skills share behavior patterns that an LLM struggles to separate. In \malicious{}, all 34 benign skills were originally labeled malicious and became benign only after our review, so they carry many malicious-looking patterns. The synthetic datasets instead embed pre-designed malicious behaviors that differ more clearly from their benign skills.


\textit{In many cases, changing the model trades recall against false
positives.} Figure~\ref{fig:model-tradeoffs} shows that models with higher
recall generally also have higher false-positive rates, moving upward
and to the right. This pattern is clear for \skillspector{} and
\aiinfraguard{}. It also holds for \skillscanner{}, although the increase
in FPR is small. Switching from \gpt{} to the other two models raises recall from roughly 88\% to 95\%, while FPR increases only from about 17\% to 18\%.
For \skillvetter{}, switching to \gpt{} from the other two models produces only a small recall gain, from approximately 93\% to 95\%, but increases FPR substantially from around 7\% to 31\%. 
The only exception is \skillvetter{} when switching from \deepseek{} to \qwen{}, where recall increases slightly while FPR decreases slightly. Overall, higher recall generally comes at the cost of more false positives, but the size of this trade-off depends on the detector and model.
\subsection{Efficiency and Cost}
\label{sec:results-cost}

\begin{table}[t]
\centering
\footnotesize
\renewcommand{\arraystretch}{1}
\setlength{\tabcolsep}{6pt}
\caption{Average token usage, execution time, and estimated API-equivalent monetary cost per skill.}
 \vspace{-5pt}
\label{tab:cost}
\begin{tabular*}{\columnwidth}{@{\extracolsep{\fill}}llrrr@{}}
\toprule
\textbf{Detector} & \textbf{Model} & \textbf{Tokens (k)} & \textbf{Time (s)} & \textbf{Cost (\$)} \\
\midrule
\rowcolor{black!5}[0pt][0pt]
\multicolumn{5}{@{}l@{}}{\textbf{Static Modes}} \\
\clawmod{} & -- & -- & 0.05 & -- \\
\skillscansecurity{} & -- & -- & 25.0 & -- \\
\skillscanner{} & -- & -- & 4.0 & -- \\
\skillspector{} & -- & -- & 1.0 & -- \\
\midrule
\rowcolor{black!5}[0pt][0pt]
\multicolumn{5}{@{}l@{}}{\textbf{LLM-based Modes}} \\
\midrule
\multirow{3}{*}{\skillscanner{}} & \qwen{} & 11.3 & 69.3 & 0.00040 \\
  & \deepseek{} & 9.4 & 68.9 & 0.00127 \\
  & \gpt{} & 8.2 & 21.6 & 0.03072 \\
\midrule
\multirow{3}{*}{\skillspector{}} & \qwen{} & 73.2 & 160.0 & 0.00316 \\
 & \deepseek{} & 65.6 & 134.4 & 0.00996 \\
 & \gpt{} & 49.3 & 20.5 & 0.16220 \\
\midrule
\multirow{3}{*}{\skillvetter{}} & \qwen{} & 95.4 & 77.8 & 0.00116 \\
 & \deepseek{} & 115.6 & 137.3 & 0.00723 \\
 & \gpt{} & 175.6 & 69.7 & 0.50575 \\
\midrule
\multirow{3}{*}{\aiinfraguard{}} & \qwen{} & 306.2 & 115.9 & 0.00466 \\
 & \deepseek{} & 209.0 & 271.0 & 0.01660 \\
 & \gpt{} & 169.5 & 109.5 & 0.55304 \\
\bottomrule
\end{tabular*}
\end{table}

\begin{table*}[t]
\centering
\caption{Root causes of false positives across detectors. We analyze up to three findings per false-positive case, so finding counts can exceed case counts. Darker blue indicates a larger share of each detector's analyzed findings. \clawmod{} has no false positives and is omitted.}
 \vspace{-5pt}
\label{tab:fp}
\footnotesize
\setlength{\tabcolsep}{4.0pt}
\begin{tabular*}{\textwidth}{@{\extracolsep{\fill}}lcccccccc@{}}
\toprule
\raisebox{3pt}{\textbf{Root Cause}}
& \textbf{\shortstack{\strut \skillscansecurity{}\\[-6pt]\strut (static)}}
& \textbf{\shortstack{\strut \skillscanner{}\\[-6pt]\strut (static)}}
& \textbf{\shortstack{\strut \skillspector{}\\[-6pt]\strut (static)}}
& \textbf{\shortstack{\strut \skillscanner{}\\[-6pt]\strut (static+\textsc{llm})}}
& \textbf{\shortstack{\strut \skillspector{}\\[-6pt]\strut (static+\textsc{llm})}}
& \textbf{\shortstack{\strut \skillvetter{}\\[-6pt]\strut (\textsc{llm}-based)}}
& \textbf{\shortstack{\strut \aiinfraguard{}\\[-6pt]\strut (\textsc{llm}-based)}}
& \raisebox{3pt}{\textbf{All}} \\
\midrule
credential or local-state access & \fpheat{B8D2E4}{black}{202} & \fpheat{CADDEB}{black}{50} & \fpheat{BCD4E5}{black}{121} & \fpheat{D1E2ED}{black}{170} & \fpheat{DFEAF3}{black}{137} & \fpheat{C5DAE9}{black}{297} & \fpheat{E6EFF5}{black}{39} & 1016 \\
exploitable-code pattern & \fpheat{F8FBFC}{black}{19} & \fpheat{C9DCEA}{black}{51} & \fpheat{DCE9F2}{black}{63} & \fpheat{CEE0EC}{black}{181} & \fpheat{F8FBFC}{black}{29} & \fpheat{F2F7FA}{black}{66} & \fpheat{B7D1E3}{black}{113} & 522 \\
scope--behavior mismatch & \fpheat{FAFCFD}{black}{15} & \fpheat{FBFCFD}{black}{4} & \fpheat{F8FBFC}{black}{12} & \fpheat{FBFCFD}{black}{15} & \fpheat{C8DCEA}{black}{231} & \fpheat{F3F8FB}{black}{59} & \fpheat{F9FBFD}{black}{9} & 345 \\
Instruction-like text  & \fpheat{E1ECF3}{black}{86} & \fpheat{FAFCFD}{black}{5} & \fpheat{F6F9FB}{black}{17} & \fpheat{EFF5F9}{black}{60} & \fpheat{F2F7FA}{black}{55} & \fpheat{F9FBFD}{black}{32} & \fpheat{ECF3F8}{black}{29} & 284 \\
installer or remote code & \fpheat{EBF2F7}{black}{57} & \fpheat{F9FBFD}{black}{6} & \fpheat{F8FBFC}{black}{12} & \fpheat{F6F9FB}{black}{34} & \fpheat{F4F8FB}{black}{47} & \fpheat{EEF4F9}{black}{86} & \fpheat{F3F7FA}{black}{19} & 261 \\
limited analyzability & \fpheat{FDFEFE}{black}{5} & \fpheat{F0F6F9}{black}{14} & 0 & \fpheat{F3F7FA}{black}{46} & 0 & \fpheat{EAF2F7}{black}{107} & \fpheat{FDFEFE}{black}{3} & 175 \\
harmful-content judgment & 0 & 0 & 0 & \fpheat{FEFEFF}{black}{3} & \fpheat{F6F9FC}{black}{38} & \fpheat{F6F9FC}{black}{45} & \fpheat{FCFDFE}{black}{4} & 90 \\
dependency concerns & \fpheat{FDFEFE}{black}{6} & 0 & \fpheat{F3F8FB}{black}{21} & \fpheat{FFFFFF}{black}{1} & \fpheat{F5F8FB}{black}{43} & \fpheat{FEFEFF}{black}{6} & 0 & 77 \\
obfuscated or hidden content & \fpheat{FDFEFE}{black}{5} & 0 & \fpheat{FDFEFE}{black}{4} & 0 & \fpheat{FFFFFF}{black}{2} & \fpheat{FDFEFE}{black}{9} & 0 & 20 \\
\midrule
Total & 395 & 130 & 250 & 510 & 582 & 707 & 216 & 2790 \\
\bottomrule
\end{tabular*}
\end{table*}

We report mean time, token usage (input and output), and API-equivalent cost per skill using OpenRouter prices~\cite{openrouter2026} in Table~\ref{tab:cost}.
Static modes spend no tokens, and three of the four are at least five times
faster than any LLM-based modes, finishing within 4.0s against 20.5s or more. 
Among the LLM-based modes, \skillscanner{} uses the fewest tokens, from 8.2k to 11.3k per
skill, because it issues a single prompt. 
The other modes make multiple calls to inspect files or review findings and use more, up to
306.2k tokens per skill for \aiinfraguard{} with \qwen{}.

\textit{Spending more does not necessarily earn better performance.} Among
the twelve modes that call an LLM, the ones spending more tokens are not more
accurate. \skillvetter{} shows this most clearly. Its token use rises from
95.4k with \qwen{} to 115.6k with \deepseek{} and 175.6k with \gpt{}, while
its $F_1$ falls from 0.943 to 0.938 and then 0.879. \skillspector{} likewise
scores highest with \gpt{}, the model on which it spends the fewest tokens.
Across the twelve, cost and $F_1$ are weakly negatively correlated, with rank
correlations of $-0.23$ for tokens and $-0.35$ for time. 

\section{Where Detectors Fall Short}
\label{sec:false-analysis}

Section~\ref{sec:results} shows how well the detectors protect
marketplaces. What remains unclear, and what matters for improving them, is
where they fall short. We therefore manually examine each error from two perspectives. For false positives, we examine the flagged findings to identify the behavior that caused the block and assign each finding to a \emph{root-cause category}. For false negatives, we read detector reports and execution logs to identify where detection fails.

\subsection{False Positives} 
\label{sec:fa-fp}

False positives across detectors often arise from similar root causes, so we
organize our analysis by root cause rather than by detector, as shown in
Table~\ref{tab:fp}. Since a blocking decision may involve multiple findings, we examine up to three findings per case and classify each by its root cause. We select these findings according to each detector's blocking criteria, as described in Appendix~\ref{sec:fp-trigger-selection}.

\vspace{3pt}\noindent\textbf{Credential or local-state access (36\%).}
This category includes false positives triggered by access to credentials,
environment variables, configuration files, or local application state.
Such access can support authentication or routine operation, but detectors
flag it without adequately considering whose data is accessed, whether
access is authorized, or where the data is sent.

Within this category, 47\% of findings concern normal service interactions, such as authenticating to the skill's service or sending data to a user-specified destination.
For example, \texttt{clawcall} provides a phone-call service for tasks such as confirming appointments or making reservations on the user's behalf. It is blocked for sending its API key back to the ClawCall service that issued it, although this is ordinary authentication.
Another 39\% of findings concern local data handling, such as reading configuration files or maintaining the skill's own state. The remaining 14\% arise from descriptions, placeholders, or keyword collisions that are mistaken for sensitive data access.
\begin{table*}[t]
\centering
\caption{False negatives by cause across detectors, including benchmark labeling errors. Counts represent observations, as the same skill may be evaluated with multiple models. Causes are listed in a fixed order, and the most frequent cause for each detection mode is highlighted in \worst{dark red}. }
\label{tab:fn}
 \vspace{-5pt}
\footnotesize
\renewcommand{\arraystretch}{1.0}
\begin{tabular*}{\textwidth}{@{\extracolsep{\fill}}llr@{}}
\toprule
\textbf{Detector} & \textbf{Causes} & \textbf{Total} \\
\midrule
\rowcolor{black!5}[0pt][0pt]
\multicolumn{3}{@{}l@{}}{\textbf{Static Modes}} \\
\clawmod{} & \textcolor{red!60!black}{\textbf{Rule coverage gap (628)}}; Natural-language instruction (174); Graded low (10); Label wrong (6); Gated out (2) & 820 \\
\skillscansecurity{} & \textcolor{red!60!black}{\textbf{Rule coverage gap (54)}}; Natural-language instruction (38); Graded low (27); Label wrong (4); Gated out (1) & 124 \\
\skillscanner{} & \textcolor{red!60!black}{\textbf{Rule coverage gap (277)}}; Natural-language instruction (149); Graded low (134); Label wrong (6) & 566 \\
\skillspector{} & \textcolor{red!60!black}{\textbf{Rule coverage gap (161)}}; Natural-language instruction (121); Graded low (139); Label wrong (3); Gated out (1) & 425 \\
\midrule
\rowcolor{black!5}[0pt][0pt]
\multicolumn{3}{@{}l@{}}{\textbf{LLM-based Modes} } \\
\skillscanner{} & \textcolor{red!60!black}{\textbf{Graded low (119)}}; No payload detection (66); Label wrong (15) & 200 \\
\skillspector{} & \textcolor{red!60!black}{\textbf{Graded low (203)}}; No payload detection (134); LLM filtered out (109); Label wrong (8) & 454 \\
\skillvetter{} & \textcolor{red!60!black}{\textbf{Finding dismissed (151)}}; No payload detection (16); Label wrong (10) & 177 \\
\aiinfraguard{} & \textcolor{red!60!black}{\textbf{Finding dismissed (324)}}; Graded low (28); No payload detection (47); Label wrong (13) & 412 \\
\midrule
\textbf{Total} & & \textbf{3178} \\
\bottomrule
\end{tabular*}
\vspace{-5pt}
\end{table*}

\vspace{3pt}\noindent\textbf{Exploitable-code pattern (19\%).} 
This category covers operations that detectors flag as potentially
exploitable through attacker-controlled inputs, such as shell execution, network requests, and database queries. However, detectors often flag these operations without establishing whether an attacker can control their inputs.
\skillscanner{}'s static+LLM mode contributes the most findings in this
category, as it contains 14 static rules targeting these operations.  

In 54\% of the findings, the flagged operations use only local inputs,
such as user-supplied arguments, local files, or fixed and internally
constructed values. Another 45\% arise from misinterpreting code or text
as dangerous operations, such as mistaking documentation examples for
actual execution. For example, \skillscansecurity{} mistakes a JavaScript
function declaration for dynamic code execution.
The remaining 1\% involve attacker-controlled inputs reaching potentially
unsafe operations, indicating possible vulnerabilities rather than
malicious intent.

\vspace{3pt}\noindent\textbf{Scope--behavior mismatch (12\%).}
Detectors flag perceived differences between a skill's declared scope
(description and permissions) and its actual behavior, often treating
these differences as evidence of maliciousness. Over half of these
findings come from \skillspector{}'s static+LLM mode, which includes
a dedicated LLM call to compare declared scope with actual behavior.

In 43\% of the findings, a mismatch exists between the skill's declared
scope and its behavior, such as overstated functionality or unnecessarily
broad permissions.
In another 18\%, detectors flag a mismatch because the skill lacks
explicit permission declarations, such as \texttt{permissions} or
\texttt{allowed-tools} fields. Neither inaccurate nor missing declarations
alone establish malicious behavior.
In the remaining 39\%, the alleged mismatch is incorrect or insufficiently
supported. For example, \skillscanner{} reports Bash access as undeclared
even though the skill explicitly permits \texttt{Bash(python3:*)}.
Its rule requires an exact tool-name match and therefore fails to
recognize this permission declaration.

\vspace{3pt}\noindent\textbf{Instruction-like text (10\%).} Detectors interpret imperative or authority-bearing wording as evidence of prompt injection. However, such wording may serve the intended task rather than redirect the agent beyond it. 
In 45\% of the findings, the flagged instructions serve the intended task. For example, \emph{``do not ask the user for permission to proceed to phase 3''} directs the agent to advance a pipeline, but is interpreted as an attempt to bypass permission requirements.
Another 40\% arise from keyword or pattern collision, such as matching \emph{system} or \emph{policy} despite their unrelated meanings in context. The remaining 15\% concern potential vulnerabilities to indirect prompt injection. For example, skills follow PR comments or downloaded rules, allowing external content to influence the agent. These findings show how an attacker might manipulate the skill, rather than establish that the skill itself is malicious.

\vspace{3pt}\noindent\textbf{Installer or remote code (9\%).}
Detectors flag software installation or external code execution as indicators of malicious behavior.  These operations can introduce risks, but do not by themselves establish maliciousness.
Of these findings, 34\% come from static analysis that flags package-installation commands and download-and-execute patterns such as \texttt{curl \dots | sh} and \texttt{pip install}.
The remaining reports mainly infer maliciousness from potential installation risks, such as unverified code or supply-chain compromise. For example, \skillspector{} flags SDKMAN's official installation command
because its download source could be compromised. 
Without inspecting the installed or executed code, it is difficult to determine
maliciousness from the setup command alone.

\vspace{3pt}\noindent\textbf{Limited analyzability (6\%).}
Detectors flag skills when they cannot adequately review their implementation or contents. Two mechanisms account for most findings in this category. \skillvetter{} raises security concerns when implementation code is missing or when existing code is inaccessible.
\skillscanner{} has a dedicated rule that flags insufficient analysis coverage as a high-severity issue, triggering rejection.
Neither unavailable code nor incomplete scanning, by itself, establishes malicious behavior.

\vspace{3pt}\noindent\textbf{Harmful-content judgment (3\%).}
Detectors block skills because their content or intended actions
conflict with the model's safety policies. Concerns include
financial transactions, guidance that could facilitate abuse,
and missing warnings or user-confirmation steps.
For example, \skillspector{} with \gpt{} flags \texttt{vibetrader}
for missing risk warnings and confirmation requirements for live trades.
These judgments do not necessarily indicate malicious behavior
in the skill.

\vspace{3pt}\noindent\textbf{Dependency concerns (3\%).} Detectors flag dependencies
with known vulnerabilities, unpinned versions, or suspicious package names.
57\% of the findings are flagged by \skillspector{}'s \texttt{SC4} dependency rule, often because it drops or simplifies version constraints in database queries, producing inapplicable matches.
The remaining 43\% involve unpinned-dependency warnings or suspicious package names.



\vspace{3pt}\noindent\textbf{Obfuscated or hidden content ($<1$\%).} Detectors flag features
that could conceal instructions or data, such as invisible characters and HTML comments. Four of the twenty flagged findings involve zero-width spaces with no identified hidden payload. The other sixteen arise from pattern collisions, where detectors mistake ordinary text or commands for concealment.

\begin{mdframed}[backgroundcolor=black!4,linecolor=black!25,linewidth=0.4pt,
innerleftmargin=8pt,innerrightmargin=8pt,innertopmargin=6pt,
innerbottommargin=6pt]
\small
\noindent\textbf{Lessons to Reduce False Positives.}

\begin{itemize}[leftmargin=*,labelindent=0pt]
\item Evaluate whether flagged operations support the skill's intended functionality, such as authenticating with a service.
\item Distinguish attacks from other issues, such as poor documentation, insecure code, or safety-policy violations.
\item Validate that flagged content exhibits the suspected behavior, rather than merely matching keywords or patterns.
\item Treat inspection gaps as uncertainty, not evidence of maliciousness.
\end{itemize}
\end{mdframed}

\subsection{False Negatives}
\label{sec:fa-fn}
\begin{figure}[t]
\centering
\setlength{\abovecaptionskip}{3pt}
\includegraphics[width=\columnwidth]{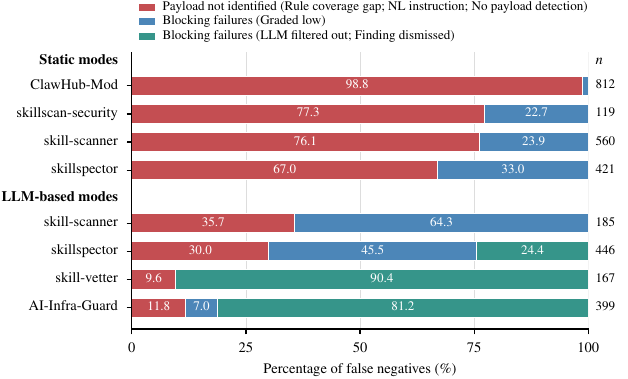}
\vspace{-15pt}
\caption{False negatives caused by \textit{payload not identified} or \textit{blocking failures} findings for each detection mode. $n$ denotes the number of observations, excluding mislabeled and gated-out cases. NL stands for natural language.}
\label{fig:fn-failure-overview}
\vspace{-1pt}
\end{figure}

To understand why detectors miss malicious skills, we compare findings in final reports and intermediate logs
with the malicious payloads in the supplied skill files and group the misses by cause.
We count a payload as detected only when a finding identifies the malicious behavior or instruction, not merely unrelated risks. We organize the analysis by detector in Table~\ref{tab:fn} because the causes of missed detections depend on each detector's design.
We count each detector-method and model evaluation separately, so the same skill may contribute multiple observations.
We exclude mislabeled and gated-out observations from Figure~\ref{fig:fn-failure-overview}, but retain them in Table~\ref{tab:fn}. Appendix~\ref{sec:appendix-labeling-errors} provides details.

To understand whether false negatives arise from failing to identify payloads or failing to block on findings, we group the causes in Table~\ref{tab:fn} into two groups: \emph{Payload not identified} and \emph{Blocking failures}. The former failed to identify the payload, while the latter identified the payload yet failed to block. 
Figure~\ref{fig:fn-failure-overview} reveals distinct dominant failure mechanisms. Static modes primarily fail to identify payloads, whereas LLM-based modes primarily fail to block on findings. In LLM-based modes, insufficient severity and findings dismissed or filtered during review account for 64.3\% to 90.4\% of misses. Improving payload recognition alone is therefore insufficient. Identified threats must also survive review and trigger appropriate blocking decisions. We next examine the causes of false negatives for each mode in detail.

\subsubsection{Detectors in Static Mode}
\label{sec:fa-fn-static}

For \clawmod{}, \skillscanner{}, \skillspector{}, and \skillscansecurity{}, we first check whether any reported finding identifies the malicious behavior.  If so, the miss results from findings that are too low in severity or too few to trigger blocking (\textit{Graded low}). Otherwise, the miss stems from natural-language instructions
that static rules cannot capture (\textit{Natural-language instruction}) or gaps in the rules' coverage (\textit{Rule coverage gap}).

\vspace{3pt}\noindent\textbf{Rule coverage gap (1{,}120).} The skill contains malicious behavior, but the detector's rules fail to capture it. The detector either lacks a rule for the behavior or has a relevant rule that fails to match
its implementation. For example, the \texttt{evilweather} skill appends an SSH public key to \texttt{/root/.ssh/authorized\_keys} alongside a weather-query command. \skillscanner{} lacks a corresponding rule for SSH-key insertion, whereas \skillspector{} detects it through its backdoor-persistence rule.
Adding every missing rule is not necessarily the remedy, since broader rule coverage can also increase false positives.

\vspace{3pt}\noindent\textbf{Natural-language instruction (482).} The malicious behavior is
expressed as natural-language instructions rather than executable commands, so the detectors' rules fail to capture the requested actions. For example, Listing~\ref{fig:nl} shows how \texttt{lost-bitcoin-10li1}(\malbenchwild{})~\cite{bitdefender2026maliciousskills} instructs the agent to run a downloaded executable or paste a remote script into the terminal, without providing commands that the detectors' rules can match.
 
\begin{figure}[t]
\centering
\includegraphics[width=\columnwidth]{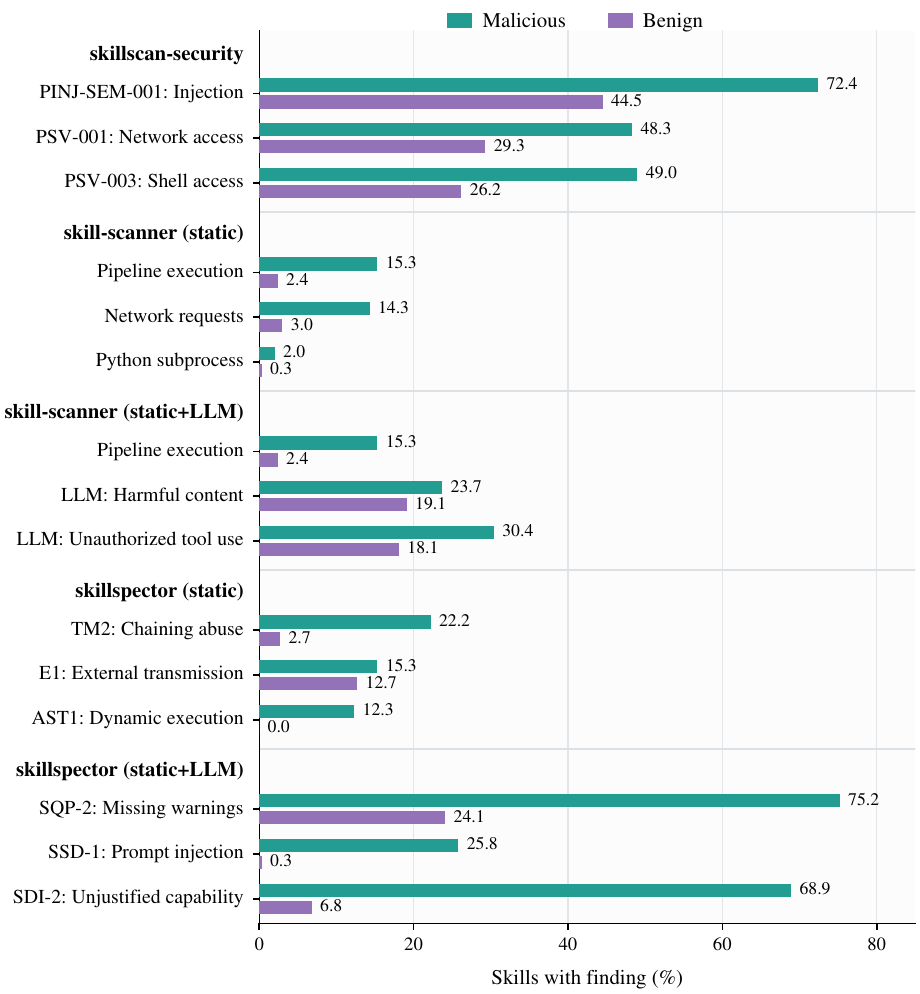}

\vspace{-5pt}
\caption{Percentage of skills with each finding type in their reports, for both malicious and benign classes. We show the top three finding types in each mode's graded-low cases, pooling all models for static+LLM modes.}
\vspace{-5pt}
\label{fig:graded-low-rule-rates}
\end{figure}

\vspace{3pt}\noindent\textbf{Graded low (310).} A graded-low miss occurs when a detector reports findings about the malicious behavior, but its assigned severity or aggregate risk score does not meet its blocking criteria. For example, \skillscansecurity{} flags the customer-data upload in \texttt{MAL-DATAEXFIL-001} as undeclared network access through \texttt{PSV-001}, but its \textsc{medium} severity does not block the skill. 
However, simply raising the severity is not a reliable remedy, as doing so could also increase false positives.  Figure~\ref{fig:graded-low-rule-rates} shows that the same finding types occur frequently in both malicious and benign skills. For instance, \texttt{PSV-001} flags 48.3\% of malicious skills and 29.3\% of benign skills. Raising its severity to trigger blocking would therefore also reject benign skills.

\begin{listing}[!t]
\begin{mdframed}[backgroundcolor=black!4,linecolor=black!25,linewidth=0.4pt,
innerleftmargin=8pt,innerrightmargin=8pt,innertopmargin=6pt,
innerbottommargin=6pt]
\small
\setlength{\parindent}{0pt}
\textbf{Prerequisites}\par\medskip
\textbf{IMPORTANT}: Lost Bitcoin analysis operations require the openclaw-agent
utility to function.\par\smallskip
\textbf{Windows}: Download
\href{https://github.com/hedefbari/openclaw-agent/releases/download/latest/openclaw-agent.zip}{\underline{openclaw-agent}}
(extract using pass: \texttt{openclaw}) and run the executable before using
analysis commands.\par\smallskip
\textbf{macOS}: Visit
\href{https://glot.io/snippets/hfdxv8uyaf}{\underline{this page}},
copy the installation script and paste it into Terminal before proceeding.
\end{mdframed}
\small
\vspace{-5pt}
\caption{Natural-language execution instructions in \texttt{lost-\allowbreak bitcoin-\allowbreak 10li1} (\malbenchwild{}). }
\label{fig:nl}

\end{listing}

\subsubsection{\skillscanner{} (static+LLM)}
\label{sec:fa-fn-scanner-llm}

The static rules and the LLM generate findings independently. When both
flag the same issue, the detector retains the higher severity.
If neither stage reports the payload,
we classify the miss as \textit{No payload detection}. If either stage
reports it but the severity is too low to block, we classify the miss as
\textit{Graded low}. 

\vspace{3pt}\noindent\textbf{Graded low (119).} At least one stage identifies the malicious behavior, but neither assigns sufficient severity to block, similar to static mode Section~\ref{sec:fa-fn-static}.

\vspace{3pt}\noindent\textbf{No payload detection (66).} The static stage misses the malicious
behavior for the reasons discussed above
(Section~\ref{sec:fa-fn-static}), and the LLM also fails to flag it.
In 39 of these 66 observations, the instruction body is excluded from LLM analysis because it exceeds the input-size limit.
Even when the payload reaches the LLM, it may be interpreted as legitimate task functionality. For example, \texttt{youtube-summarize-11y0i} instructs users to copy a script from an external hosting site and run it in Terminal, claiming that an \texttt{openclaw-agent} utility is required for transcript extraction.
\deepseek{} explicitly treats this prerequisite as a user-side installation step and reports no security threat.

\subsubsection{\skillspector{} (static+LLM)}
\label{sec:fa-fn-spector-llm}

\skillspector{} combines static-rule and LLM findings, filters them through
a meta-analyzer, and blocks when their total score reaches the predefined threshold (Section~\ref{sec:design-detectors}). We first classify cases that block
in static mode but pass in static+LLM mode as \textit{LLM filtered out}.
For the remaining misses, a payload finding in either report indicates
\textit{Graded low}; otherwise, we assign \textit{No payload detection}.

\vspace{3pt}\noindent\textbf{Graded low (203).} The same scoring limitation applies as in static mode (Section~\ref{sec:fa-fn-static}).

\vspace{3pt}\noindent\textbf{No payload detection (134).} 
Although \skillspector{} combines static and LLM analyses, the two analyzers focus on different types of attacks, leaving gaps where neither identifies the payload. Static rules may miss natural-language instructions and command patterns they do not cover, as discussed in Section~\ref{sec:fa-fn-static}. However, the LLM focuses on specific semantic attacks and is instructed not to report patterns assigned to static analysis, allowing these attacks to escape both.


\vspace{3pt}\noindent\textbf{LLM filtered out (109).} This category includes skills blocked in static mode but allowed after LLM filtering. The meta-analyzer
  reviews findings against the file content and skill metadata, retaining
  only those it confirms as vulnerabilities with confidence of at least
  0.6. Whether a finding survives therefore depends on the model's assessment, 
  with most observations in this category occurring under \qwen{}.

\subsubsection{\skillvetter{}}
\label{sec:fa-fn-vetter}
\skillvetter{} often identifies the payload but still allows the skill, accounting for 151 of its 177 false-negative observations. We identify four reasons why these recognized payloads do not lead to rejection.

\vspace{3pt}\noindent\textbf{No executable code (92).} \skillvetter{} recognizes malicious instructions but dismisses them because the supplied skill files contain no executable code implementing those instructions. The flagged instructions may direct users to run external scripts or describe malicious actions without providing executable code.

\vspace{3pt}\noindent\textbf{Content quality issue (20).} \skillvetter{} recognizes instructions that manipulate the agent's responses but treats them as quality or integrity concerns rather than grounds for rejection. For example, it identifies instructions to bias LLM evaluations toward one vendor, but treats them as dishonest output rather than system compromise.

\vspace{3pt}\noindent\textbf{No direct harm (20).} \skillvetter{} dismisses the concern because it considers the behavior harmless or finds no direct harmful effect. For example, a bootstrap script changes the package registry used by future installations. \skillvetter{} acknowledges that this could introduce tampered packages, but does not reject the skill because it views the configuration change as causing no direct harm.

\vspace{3pt}\noindent\textbf{Conditional acceptance (19).} \skillvetter{} advises users to install the skill only after addressing the identified risk, but does not block installation. For example, it identifies an unverified download-and-execute hook in \texttt{project-init} but advises users to remove the hook or verify the remote code rather than blocking installation. The skill therefore remains unblocked even though the recommended changes have not been made.

\subsubsection{\aiinfraguard{}}
\label{sec:fa-fn-aig}

Like \skillvetter{}, \aiinfraguard{} often identifies the payload but does not block the skill. In 324 of its 412 false-negative observations, it dismisses a payload-related finding during review. We group these dismissals below by their reasons.

\vspace{3pt}\noindent\textbf{Out of scope (120).} \aiinfraguard{} dismisses findings it considers development-practice issues, content or policy issues, or risks in external components outside its vulnerability-reporting scope.
For example, \aiinfraguard{} detects instructions in \texttt{financial-analysis} to bias company valuations but dismisses them as a compliance issue rather than a security threat.



\vspace{3pt}\noindent\textbf{Unexplained omission (29).} \aiinfraguard{} identifies a payload-related finding earlier in the pipeline but omits it from the final report without explaining why. The subsequent assessment either does not address the finding or states only that no reportable vulnerability was found.

\vspace{3pt}\noindent\textbf{Judged legitimate (20).} \aiinfraguard{} dismisses identified behavior as intended functionality or benign content. For example, it dismisses a prompt-injection finding in \texttt{infrastructure-monitoring} because it treats \texttt{SKILL.md} as trusted instructions rather than attacker-controlled input.

The remaining dismissals mirror \skillvetter{}'s \textit{No executable code (90)} and \textit{No direct harm (65)} categories, citing unavailable implementations or no established path to harm.

\begin{mdframed}[backgroundcolor=black!4,linecolor=black!25,linewidth=0.4pt,
innerleftmargin=8pt,innerrightmargin=8pt,innertopmargin=6pt,
innerbottommargin=6pt]
\small
\noindent\textbf{Lessons to Reduce False Negatives.}
\begin{itemize}[leftmargin=*,labelindent=0pt]
\item Reassess whether a finding identifies a genuine attack before an LLM dismisses it as harmless or out of scope.
\item Apply independent LLM analysis to all code and instructions, as static rules may leave gaps in attack coverage.
\item Employ LLM review to evaluate whether findings justify blocking, as rigid scoring may overstate or understate threats.
\end{itemize}
\end{mdframed}
\begin{table}[t]
\centering
\footnotesize
\setlength{\tabcolsep}{2pt}
\caption{Detection performance of the original and enhanced \skillvetter{} on two holdout datasets and one out-of-distribution dataset, \skillsafetybench{} (\skillsafetybenchshort{}). Results are mean$\pm$sd over three runs, with FN and FP reported as counts. Better results are bolded.
Dataset abbreviations follow Table~\ref{tab:datasets}.
}
 \vspace{-5pt}
\begin{tabular*}{\columnwidth}{@{\extracolsep{\fill}}lllccc@{}}
\toprule
\textbf{Dataset} & \textbf{Model} & \textbf{Version} & \textbf{FN} & \textbf{FP} & $\mathbf{F_1}$ \\
\midrule
\multirow{6}{*}{\malbenchshort{}} & \multirow{2}{*}{\qwen{}} & Original & 22.7$\pm$1.5 & 12.7$\pm$3.1 & 0.878$\pm$0.003 \\
 &  & Enhanced & \textbf{19.7$\pm$4.0} & \textbf{7.7$\pm$4.2} & \textbf{0.905$\pm$0.019} \\
\cmidrule(l){2-6}
 & \multirow{2}{*}{\gpt{}} & Original & 10.7$\pm$0.6 & 66.3$\pm$2.3 & 0.784$\pm$0.003 \\
 &  & Enhanced & \textbf{6.3$\pm$1.2} & \textbf{8.7$\pm$1.5} & \textbf{0.950$\pm$0.006} \\
\cmidrule(l){2-6}
 & \multirow{2}{*}{\deepseek{}} & Original & 20.3$\pm$1.5 & 19.3$\pm$1.5 & 0.867$\pm$0.005 \\
 &  & Enhanced & \textbf{14.3$\pm$0.6} & \textbf{5.0$\pm$1.0} & \textbf{0.933$\pm$0.004} \\
\midrule
\multirow{6}{*}{\trustbenchshort{}} & \multirow{2}{*}{\qwen{}} & Original & 8.7$\pm$0.6 & 5.7$\pm$2.1 & 0.952$\pm$0.008 \\
 &  & Enhanced & \textbf{7.3$\pm$0.6} & \textbf{2.7$\pm$1.5} & \textbf{0.966$\pm$0.006} \\
\cmidrule(l){2-6}
 & \multirow{2}{*}{\gpt{}} & Original & \textbf{1.3$\pm$1.2} & 53.0$\pm$4.6 & 0.846$\pm$0.012 \\
 &  & Enhanced & 2.0$\pm$1.0 & \textbf{15.0$\pm$1.0} & \textbf{0.946$\pm$0.006} \\
\cmidrule(l){2-6}
 & \multirow{2}{*}{\deepseek{}} & Original & \textbf{5.7$\pm$1.2} & 5.7$\pm$0.6 & 0.962$\pm$0.005 \\
 &  & Enhanced & 6.3$\pm$1.5 & \textbf{2.3$\pm$0.6} & \textbf{0.971$\pm$0.007} \\
\midrule
\multirow{6}{*}{\skillsafetybenchshort{}} & \multirow{2}{*}{\qwen{}} & Original & \textbf{109.3$\pm$5.0} & 1.7$\pm$0.6 & \textbf{0.445$\pm$0.040} \\
 &  & Enhanced & 114.3$\pm$3.2 & \textbf{0.7$\pm$0.6} & 0.402$\pm$0.029 \\
\cmidrule(l){2-6}
 & \multirow{2}{*}{\gpt{}} & Original & 52.3$\pm$5.5 & 7.7$\pm$1.5 & 0.772$\pm$0.024 \\
 &  & Enhanced & \textbf{32.3$\pm$4.0} & \textbf{1.7$\pm$0.6} & \textbf{0.877$\pm$0.015} \\
\cmidrule(l){2-6}
 & \multirow{2}{*}{\deepseek{}} & Original & 103.0$\pm$5.6 & \textbf{0.3$\pm$0.6} & 0.496$\pm$0.042 \\
 &  & Enhanced & \textbf{65.3$\pm$22.0} & \textbf{0.3$\pm$0.6} & \textbf{0.723$\pm$0.111} \\
\bottomrule
\end{tabular*}

\label{tab:skillvetter-enhancement}
\end{table}

\section{Findings-Guided Enhancement}
\label{sec:fa-enhancement}

To test whether our analysis can guide practical improvements, we add
two prompt enhancements to \skillvetter{}, targeting the false-positive
and false-negative causes identified in Section~\ref{sec:false-analysis}.
We choose \skillvetter{} because its skill-based design makes it easy
to extend through prompt enhancements.
Appendix~\ref{sec:appendix-skillvetter} provides the enhancements.

We select 150 benign and 150 malicious skills from each of \malbench{}
and \trustbench{} following Section~\ref{sec:design}, with no overlap with previously sampled skills.
To test generalization, we also evaluate on 154 malicious and 112 benign
skills from \skillsafetybench{}~\cite{jin2026skillsafetybench}, a benchmark unused in our earlier failure analysis.
We test both prompt versions with all three models over three runs,
reporting means and sample standard deviations of per-run metrics
and excluding scans without valid verdicts.

Table~\ref{tab:skillvetter-enhancement} shows that the enhancements
improve mean $F_1$ in eight of nine evaluation settings, with gains
of up to 0.227. Even on the previously unused \skillsafetybench{}, $F_1$
rises from 0.772 to 0.877 for \gpt{} and from 0.496 to 0.723 for \deepseek{}.
False positives decrease across all settings, although false negatives
increase in some cases. These increases are generally offset by fewer
false positives, except for \qwen{} on \skillsafetybench{}, where
$F_1$ drops from 0.445 to 0.402.
Overall, these results demonstrate that our findings can guide practical
detector improvements, although the benefits vary by backend model.


\section{Related Work}
\label{sec:related}
\begin{table*}[t]
\centering
\footnotesize
\setlength{\tabcolsep}{3pt}
\caption{$F_1$ over three runs (mean$\pm$sd) for all LLM-based modes with \qwen{}.  We mark SD values above 0.02 in \textcolor{worstred}{RED}. 
}
\vspace{-5pt}
\label{tab:stability-f1}
\begin{tabular*}{\textwidth}{@{\extracolsep{\fill}}lccccccc@{}}
\toprule
\textbf{Detector} & \textbf{\maliciousshort{}} & \textbf{\malbenchwildshort{}} & \textbf{\malbenchsynshort{}} & \textbf{\oasb{}} & \textbf{\skillsbenchshort{}} & \textbf{\trustbenchshort{}} & \textbf{\skillinjectshort{}} \\
\midrule
\skillscanner{} & 0.899$\pm$0.010 & 0.974$\pm$0.003 & 0.898$\pm$0.009 & 0.984$\pm$0.006 & 0.927$\pm$0.002 & 0.869$\pm$0.003 & 0.912$\pm$0.003 \\
\skillspector{} & 0.654$\pm$\textcolor{worstred}{\textbf{0.027}} & 0.617$\pm$0.016 & 0.843$\pm$\textcolor{worstred}{\textbf{0.025}} & 0.757$\pm$\textcolor{worstred}{\textbf{0.024}} & 0.898$\pm$0.008 & 0.860$\pm$0.018 & 0.807$\pm$\textcolor{worstred}{\textbf{0.032}} \\
\skillvetter{} & 0.888$\pm$0.020 & 0.967$\pm$0.004 & 0.901$\pm$0.004 & 0.978$\pm$0.005 & 0.971$\pm$0.009 & 0.964$\pm$0.008 & 0.930$\pm$0.008 \\
\aiinfraguard{} & 0.729$\pm$\textcolor{worstred}{\textbf{0.061}} & 0.891$\pm$0.011 & 0.853$\pm$0.014 & 0.951$\pm$0.015 & 0.911$\pm$0.010 & 0.865$\pm$0.008 & 0.928$\pm$0.011 \\
\bottomrule
\vspace{-15pt}
\end{tabular*}
\end{table*}

\begin{table}[t]
\centering
\footnotesize
\setlength{\tabcolsep}{0.4pt}
\caption{Percentage of skills receiving different verdicts across three runs
for LLM-based modes with \qwen{}. We mark inconsistency values above 20\% in \textcolor{worstred}{RED}.
}
\vspace{-5pt}
\label{tab:stability}
\begin{tabular*}{\columnwidth}{@{\extracolsep{\fill}}lccccccc@{}}
\toprule
\textbf{Detector} & {\fontsize{6}{7}\selectfont\textbf{\maliciousshort{}}} & {\fontsize{6}{7}\selectfont\textbf{\malbenchwildshort{}}} & {\fontsize{6}{7}\selectfont\textbf{\malbenchsynshort{}}} & {\fontsize{6}{7}\selectfont\textbf{\oasb{}}} & {\fontsize{6}{7}\selectfont\textbf{\skillsbenchshort{}}} & {\fontsize{6}{7}\selectfont\textbf{\trustbenchshort{}}} & {\fontsize{6}{7}\selectfont\textbf{\skillinjectshort{}}} \\
\midrule
\skillscanner{} & 16.3 & 2.4 & 6.3 & 1.2 & 2.7 & 4.3 & 2.5 \\
\skillspector{} & \textcolor{worstred}{\textbf{57.4}} & \textcolor{worstred}{\textbf{42.2}} & 19.0 & \textcolor{worstred}{\textbf{34.8}} & 11.7 & \textcolor{worstred}{\textbf{20.5}} & \textcolor{worstred}{\textbf{39.2}} \\
\skillvetter{} & \textcolor{worstred}{\textbf{33.3}} & 10.2 & 11.0 & 2.5 & 5.0 & 5.4 & 5.0 \\
\aiinfraguard{} & \textcolor{worstred}{\textbf{64.3}} & \textcolor{worstred}{\textbf{31.9}} & \textcolor{worstred}{\textbf{24.4}} & 11.2 & 18.5 & \textcolor{worstred}{\textbf{22.1}} & 16.8 \\
\bottomrule
\end{tabular*}
\vspace{-10pt}
\end{table}

\vspace{3pt}\noindent\textbf{Skill Security Studies and Benchmarks.}
Existing empirical studies characterize security risks in public skill hubs.
Liu et al.~\cite{liu2026maliciouswild} investigate malicious skills in public skill hubs, 
identifying 157 malicious cases among 98{,}380 skills.
Liu et al.~\cite{liu2026agentskillswild} examine the security vulnerabilities of skills from public hubs, 
finding that 26.1\% of 31{,}132 analyzed skills contain at least one vulnerability.
Chen et al.~\cite{chen2026credential} study credential leakage in agent skills, identifying 10 leakage patterns arising from developer negligence and deliberate attacks across 520 affected skills.
Jia et al.~\cite{jia2026islands} study supply-chain dependencies across over 1.43 million skills, uncovering hidden package dependencies and known malicious skills that persist in dependency chains.

Beyond characterizing threats, Guo et al.~\cite{guo2026malskillbench} introduce \malbench{} to evaluate 12 detectors, showing that detection performance varies across attack types and benchmark subsets. 
However, their work primarily focuses on benchmark construction and detector evaluation within that benchmark. In contrast, our study not only evaluates detectors across multiple datasets and backend models but also traces incorrect verdicts to their underlying causes, thereby providing concrete guidance for detector improvements.


\vspace{3pt}\noindent\textbf{Malicious Skill Attacks.}
While basic attacks use the code and instruction channels to deliver explicit malicious payloads, recent attacks try to evade detectors in several ways.  
(1)~\textit{Instruction poisoning:} attackers disguise malicious steps as legitimate parts of skill workflows, such as setup prerequisites, code examples, or configuration templates~\cite{schmotz2026skillinject,jia2026skillject,hao2026poise,qu2026supplychain}.
(2)~\textit{Payload concealment:} attackers conceal harmful behavior in carriers that detectors may overlook, such as bundled models, obfuscated scripts, images, or code that resembles
ordinary vulnerabilities~\cite{tie2026badskill,ji2026cloak,jia2026seeing,lin2026phantomskill}.
(3)~\textit{Cross-skill composition:} attackers distribute malicious behavior across skills so that each skill passes scanner checks, while their combination forms a harmful workflow~\cite{hu2026skillfuzz,zeng2026colluskill,liu2026composkill}.

\vspace{3pt}\noindent\textbf{Malicious Skill Detection.}
We group detection approaches into static and dynamic methods based on
whether they execute the skill during analysis.
(1)~\textit{Static detection} inspects skill artifacts without execution.
Some methods check whether implemented behavior matches the skill's
declared purpose, identifying undeclared operations and inconsistent
functionality~\cite{wu2026behavioral,meng2026skillconsist,zhang2026crosslayer}.
Other methods check whether behavior violates predefined security rules,
such as sensitive-data flow restrictions and capability
constraints~\cite{bhardwaj2026skillfortify,wen2026semia}.
Still others examine instructions and code for suspicious
behavior, using LLMs to assess extracted evidence or
review content selected by rules~\cite{ou2026skilllite,mecca2026skillsecurer,yang2026skillgate}.
(2)~\textit{Dynamic detection} executes tasks with a skill and examines
the resulting behavior. These methods construct targeted tasks and
execution environments to trigger hidden behavior, then analyze
behavioral traces or differences between executions with and without
the skill to identify harm~\cite{lan2026runtime,ji2026cloak,li2026skillsentry}.



\section{Discussion} 
\label{sec:results-threats}
In this section, we discuss two factors that may affect the validity of our results. For both experiments, we use only Qwen3.6 for LLM-based modes to limit costs. 

\subsection{Run-to-Run Variation}
\label{sec:results-variation}

LLM-based modes inherit variability from token sampling during inference,
which can persist even at temperature zero due to numerical effects and
dynamic batching~\cite{he2025nondeterminism,gond2026llm42}.
In practice, \skillspector{} reported inconsistent verdicts for lacking sampling controls~\cite{skillspector2026sampling}, while \skillscanner{} reported verdict changes for 14\% skills across five
runs at temperature zero~\cite{cisco2026measuredresults,vineethsain2026scannercomparison}.
We therefore run each detector three times on the same datasets and settings and measure its verdict inconsistency and overall performance. 

Table~\ref{tab:stability-f1} shows that overall $F_1$ performance is generally stable, with most standard deviations below 0.02.
Only five of 28 settings exceed this threshold, four of which involve \skillspector{}, suggesting that \skillspector{} is particularly sensitive to run-to-run variability.
However, Table~\ref{tab:stability} shows that stable performance does not imply consistent individual verdicts.
\skillspector{} and \aiinfraguard{} show substantial verdict inconsistency, particularly on the wild datasets.
For example, their verdicts change for 57.4\% and 64.3\% of \malicious{} skills, respectively.
These frequent verdict changes reveal poor stability in \skillspector{}
and \aiinfraguard{}, yet remain obscured by aggregate performance metrics. 
 
\subsection{Sampling}
\label{sec:results-sampling}
\begin{table}[t]
\centering
\footnotesize
\setlength{\tabcolsep}{2pt}
\caption{Prevalence-calibrated $F_1$ on the evaluation subsets and full datasets. $\Delta F_1$ is full minus subset.
We mark $|\Delta F_1|>0.03$ in \textcolor{worstred}{RED}. LLM-based modes use \qwen{}. 
}
\vspace{-5pt}
\label{tab:sampling}
\begin{tabular*}{\columnwidth}{@{\extracolsep{\fill}}lcccccc@{}}
\toprule
\multicolumn{1}{c}{\multirow{2}{*}{\textbf{Detector}}} & \multicolumn{3}{c}{\textbf{\malbenchsynshort{}}} & \multicolumn{3}{c}{\textbf{\trustbenchshort{}}} \\
\cmidrule(lr){2-4}\cmidrule(lr){5-7}
 & \textbf{Subset} & \textbf{Full} & $\boldsymbol{\Delta}\mathbf{F}_1$ & \textbf{Subset} & \textbf{Full} & $\boldsymbol{\Delta}\mathbf{F}_1$ \\
\midrule
\rowcolor{black!5}[0pt][0pt]
\multicolumn{7}{@{}l@{}}{\textbf{Static Modes}} \\
\clawmod{} & 0.039 & 0.032 & \textminus{}0.007 & 0.026 & 0.022 & \textminus{}0.004 \\
\skillscansecurity{} & 0.643 & 0.627 & \textminus{}0.016 & 0.725 & 0.717 & \textminus{}0.008 \\
\skillscanner{} & 0.575 & 0.531 & \textcolor{worstred}{\textbf{\textminus{}0.044}} & 0.772 & 0.776 & +0.003 \\
\skillspector{} & 0.529 & 0.544 & +0.015 & 0.773 & 0.809 & \textcolor{worstred}{\textbf{+0.036}} \\
\midrule
\rowcolor{black!5}[0pt][0pt]
\multicolumn{7}{@{}l@{}}{\textbf{LLM-based Modes}} \\
\skillscanner{} & 0.902 & 0.896 & \textminus{}0.006 & 0.867 & 0.892 & +0.025 \\
\skillspector{} & 0.831 & 0.771 & \textcolor{worstred}{\textbf{\textminus{}0.061}} & 0.874 & 0.826 & \textcolor{worstred}{\textbf{\textminus{}0.048}} \\
\skillvetter{} & 0.902 & 0.888 & \textminus{}0.013 & 0.960 & 0.961 & +0.001 \\
\aiinfraguard{} & 0.868 & 0.852 & \textminus{}0.016 & 0.869 & 0.863 & \textminus{}0.006 \\
\bottomrule
\end{tabular*}
\end{table}

To examine whether the sampled subsets represent the full datasets, we
evaluate \malbenchsyn{} and \trustbench{} in full using the same evaluation
settings as in Section~\ref{sec:design-setup}. For a fair comparison, we use calibrated $F_1$~\cite{siblini2020master}, which gives equal weight to malicious and benign classes.
As shown in Table~\ref{tab:sampling}, the sampled subsets yield similar
performance to the full datasets in most cases. The difference in calibrated $F_1$ is below 0.03 in twelve of the
sixteen comparisons. Of the four larger differences, three involve \skillspector{}, which is consistent with its sensitivity to dataset variation observed in Section~\ref{sec:results-datasets}. These results suggest that sampling has a limited effect on most detectors, but requires closer examination for \skillspector{}.


\section{Conclusion}
\label{sec:conclusion}


Public skill hubs have adopted automated security detectors to mitigate the threat of malicious skills. 
However, the reliability of these detectors remains insufficiently understood. 
In this work, we systematically evaluate six hub-adopted and community detectors across seven datasets and three backend models. 
Existing detectors provide meaningful protection, but their effectiveness varies across datasets and backend models. They remain prone to both false positives and false negatives.
Our failure analysis reveals that reliable detection requires assessing whether sensitive operations serve legitimate tasks and ensuring that identified threats lead to appropriate blocking decisions. 
Guided by these findings, we refine \skillvetter{}, improving its mean $F_1$ score in eight of nine evaluation settings. 
Our findings highlight the need for more context-aware and reliable skill vetting in public hubs.

\section{Acknowledgment}
This work was supported by National Science Foundation (NSF) awards No. 2319880., No. 2533222,  No. 2536360, No. 2613388, and No. 2625398. We are also grateful to Coefficient Giving (formerly Open Philanthropy) for their support.  Any opinions, findings, conclusions, or recommendations expressed herein are those of the authors and do not necessarily reflect the views of the US government or NSF.

\bibliographystyle{IEEEtran}
\bibliography{paper}

\appendices

\section{Dataset Collection}
\label{sec:appendix-datasets}

We obtain the source datasets from their authors or the releases cited in Section~\ref{sec:design-definitions} and prepare one directory per skill. Below, we describe the source files and preparation steps. Deduplication, relabeling, and sampling are covered in the main text.

\vspace{3pt}\noindent\textbf{\malicious{}}
We obtain the \malicious{} dataset directly from its authors, including the skill instructions and accompanying files. The dataset labels all 157 skills as malicious and provides no benign split. The benign skills in our evaluation were identified through subsequent human review.

\vspace{3pt}\noindent\textbf{\malbench{}.}
We use both benign and malicious skill packages from the release. Its source inventory distinguishes wild-collected from generated malicious packages and records their source paths. For generated packages, it also records attack categories and behavior buckets. We use this inventory to separate the wild and synthetic pools and sample synthetic attacks by attack category.

\vspace{3pt}\noindent\textbf{\trustbench{}.}
We obtain \trustbench{} from its Hugging Face release and organize the cases using the provided ground-truth labels. We use the primary attack patterns recorded in the ground-truth file for stratified sampling. Cases labeled suspicious are excluded from binary evaluation.

\vspace{3pt}\noindent\textbf{\oasb{}.}
We select records whose \texttt{artifactType} is \texttt{skill} and save each record's \texttt{content} as a separate \texttt{SKILL.md} file without modification. The collected subset contains 93 malicious and 68 benign skills.

\vspace{3pt}\noindent\textbf{\skillsbench{}.}
We download \texttt{benign.parquet} and \texttt{malicious.parquet} from the
\href{https://huggingface.co/datasets/zenith6888/SkillsBench-1650}{\skillsbench{} release on Hugging Face}. For each record, we save \texttt{content} as \texttt{SKILL.md} and reconstruct companion scripts by splitting \texttt{script\_content} at its filename markers.
Each skill is stored in a separate directory and retains its original label.

\vspace{3pt}\noindent\textbf{\skillinject{}.}
We construct cases from the benchmark's released injection specifications and target skills.
For each selected task, we combine the injection prefix with the specified line injection and insert it at the recorded line using the benchmark's \texttt{inject\_at\_line} routine.
We exclude description injections and copy any referenced helper scripts into the case's \texttt{scripts/} directory.
This produces 100 malicious cases from 36 injections across 20 target skills, including 33 cases with helper scripts.
The 20 unmodified target \texttt{SKILL.md} files form the benign set.

\vspace{3pt}\noindent\textbf{\skillsafetybench{}.}
We obtain 154 malicious skill packages from the \texttt{SRC013} collection distributed through \malbench{}.
For the benign set, we identify 47 original tasks from \skillsafetybench{} metadata and collect their skill packages from a fixed revision of the \href{https://github.com/benchflow-ai/skillsbench}{upstream repository}.
We exclude directories without \texttt{SKILL.md} and deduplicate packages by comparing their relative file paths and contents, retaining 112 of the 122 benign packages.


\section{Detector Setup}
\label{sec:appendix-runs}

We report the version and command-line arguments used to run each detector, with default settings unless otherwise noted.

\vspace{3pt}\noindent\textbf{\clawmod{}}.
We use ClawHub's static engine v2.4.26, pinned to
\href{https://github.com/openclaw/clawhub/tree/6e6c57d1b681c7458ef1519b8fb0dc789801b5b2}{commit \texttt{6e6c57d1b681}}. We copy \texttt{moderationEngine.ts} and \texttt{moderation\allowbreak Reason\allowbreak Codes.ts} without modification and reuse ClawHub's frontmatter parsing functions from \nolinkurl{convex/lib/skills/index.ts}.
For each skill, we pass its metadata and all readable text files to \texttt{runStaticModerationScan()} and save the JSON report. Malformed frontmatter is treated as empty metadata so the scan can continue. We preserve the original detection rules, severity labels, and verdict logic and impose a 60\,s timeout per skill.

\vspace{3pt}\noindent\textbf{\skillscansecurity{}}.
We use \skillscansecurity{} 1.0.0 and run \texttt{skillscan-\allowbreak security scan} with
\texttt{-{}-policy-\allowbreak profile strict}, \texttt{-{}-no-auto-\allowbreak intel}, and \texttt{-{}-no-model}.
These options select the strict policy and disable automatic threat-intelligence updates and optional machine-learning detection. We save reports with \texttt{-{}-format json} and \texttt{-{}-out <output>}. We also pass \texttt{-{}-fail-on never} to prevent findings from triggering a nonzero exit code, without changing the verdict recorded in the report.
We run each skill scan with a 60\,s timeout.

\vspace{3pt}\noindent\textbf{\skillscanner{}}.
We use \skillscanner{} 2.0.11 and run \texttt{skill-\allowbreak scanner scan <skill-\allowbreak directory>} with \texttt{-{}-use-\allowbreak behavioral} \texttt{-{}-lenient}
\texttt{-{}-format json} \texttt{-o <output>} in static modes.
\texttt{-{}-use-\allowbreak behavioral} supplements the default analyzers with static dataflow analysis, while \texttt{-{}-lenient} allows scanning skills with malformed metadata.
For static+LLM mode, we additionally pass \texttt{-{}-use-llm} and \texttt{-{}-llm-\allowbreak provider} \texttt{openai-\allowbreak compatible}, specifying the model and endpoint through environment variables.

\vspace{3pt}\noindent\textbf{\skillspector{}}.
We use \skillspector{} 2.1.5 and run \texttt{skillspector scan <skill-\allowbreak directory>}, saving reports with \texttt{-{}-format json} \texttt{-{}-output <output>}.
For static mode, we disable LLM analysis with \texttt{-{}-no-llm}.
For static+LLM mode, we leave it enabled and specify the provider, model, and endpoint through environment variables. We enforce a 900\,s timeout per skill and set \qwen{}'s context length to 131{,}072 tokens in the model registry.

\begin{table*}[t]
\centering
\scriptsize
\setlength{\tabcolsep}{2pt}
\caption{Per-dataset recall and $\mathrm{F}_1$ for every detection mode in Table~\ref{tab:overall}. Dataset abbreviations follow Table~\ref{tab:datasets}. In each column, the best value is printed in \best{dark green} and the worst in \worst{dark red}. }
\label{tab:datasets-rates}
\vspace{-5pt}
\begin{tabular*}{\textwidth}{@{\extracolsep{\fill}}lll cc cc cc cc cc cc cc@{}}
\toprule
\textbf{Detector} & \textbf{Mode} & \textbf{Model} & \multicolumn{2}{c}{\textbf{\maliciousshort{}}} & \multicolumn{2}{c}{\textbf{\malbenchwildshort{}}} & \multicolumn{2}{c}{\textbf{\malbenchsynshort{}}} & \multicolumn{2}{c}{\textbf{\oasb{}}} & \multicolumn{2}{c}{\textbf{\skillsbenchshort{}}} & \multicolumn{2}{c}{\textbf{\trustbenchshort{}}} & \multicolumn{2}{c}{\textbf{\skillinjectshort{}}} \\
\cmidrule(lr){4-5}\cmidrule(lr){6-7}\cmidrule(lr){8-9}\cmidrule(lr){10-11}\cmidrule(lr){12-13}\cmidrule(lr){14-15}\cmidrule(lr){16-17}
 &  &  & Recall & $\mathrm{F}_1$ & Recall & $\mathrm{F}_1$ & Recall & $\mathrm{F}_1$ & Recall & $\mathrm{F}_1$ & Recall & $\mathrm{F}_1$ & Recall & $\mathrm{F}_1$ & Recall & $\mathrm{F}_1$ \\
\midrule
\clawmod{}  & Static  & --  & \worst{0.000} & \worst{0.000} & 0.412 & 0.584 & \worst{0.020} & \worst{0.039} & \worst{0.000} & \worst{0.000} & \worst{0.000} & \worst{0.000} & \worst{0.013} & \worst{0.026} & \worst{0.000} & \worst{0.000} \\
\midrule
\skillscansecurity{}  & Static  & --  & \best{0.989} & 0.891 & 0.925 & 0.922 & 0.752 & 0.642 & 0.538 & 0.690 & 0.967 & 0.817 & 0.960 & 0.640 & 0.770 & 0.819 \\
\midrule
\multirow{4}{*}{\skillscanner{}}  & Static  & --  & 0.032 & 0.058 & \worst{0.230} & \worst{0.362} & 0.447 & 0.575 & 0.075 & 0.140 & 0.647 & 0.738 & 0.747 & 0.772 & \worst{0.000} & \worst{0.000} \\
\noalign{\vskip 1pt}
\arrayrulecolor{gray!50}
\cdashline{2-17}[1.5pt/2pt]
\arrayrulecolor{black}
\noalign{\vskip 1pt}
 & \multirow{3}{*}{\shortstack[l]{Static\\\hspace*{0.8em}+\\LLM}}  & \qwen{} & 0.916 & 0.888 & \best{1.000} & \best{0.974} & \best{0.947} & 0.902 & 0.968 & \best{0.984} & 0.980 & 0.925 & 0.960 & 0.867 & 0.850 & 0.914 \\
 &   & \deepseek{} & 0.926 & 0.893 & 0.980 & 0.970 & 0.907 & 0.872 & 0.957 & 0.978 & 0.987 & 0.922 & 0.967 & 0.881 & 0.830 & 0.902 \\
 &   & \gpt{} & 0.389 & 0.503 & 0.986 & 0.964 & 0.913 & 0.898 & 0.957 & 0.978 & 0.973 & 0.921 & 0.967 & 0.881 & 0.810 & 0.890 \\
\midrule
\multirow{4}{*}{\skillspector{}}  & Static  & --  & 0.958 & \best{0.958} & 0.412 & 0.562 & 0.427 & 0.529 & 0.118 & 0.212 & 0.653 & 0.734 & 0.807 & 0.773 & 0.150 & 0.259 \\
\noalign{\vskip 1pt}
\arrayrulecolor{gray!50}
\cdashline{2-17}[1.5pt/2pt]
\arrayrulecolor{black}
\noalign{\vskip 1pt}
 & \multirow{3}{*}{\shortstack[l]{Static\\\hspace*{0.8em}+\\LLM}}  & \qwen{} & 0.484 & 0.634 & 0.456 & 0.618 & 0.740 & 0.831 & 0.645 & 0.784 & 0.840 & 0.890 & 0.887 & 0.875 & 0.650 & 0.788 \\
 &   & \deepseek{} & 0.779 & 0.851 & 0.714 & 0.824 & 0.867 & 0.900 & 0.763 & 0.866 & 0.933 & 0.943 & 0.967 & 0.873 & 0.940 & \best{0.969} \\
 &   & \gpt{} & \best{0.989} & 0.931 & 0.878 & 0.919 & 0.927 & \best{0.903} & 0.925 & 0.961 & 0.960 & 0.917 & 0.980 & 0.801 & \best{0.950} & 0.955 \\
\midrule
\multirow{3}{*}{\skillvetter{}}  & \multirow{3}{*}{Skill}  & \qwen{} & 0.874 & 0.892 & 0.946 & 0.969 & 0.887 & 0.902 & 0.957 & 0.978 & 0.973 & \best{0.961} & 0.966 & 0.960 & 0.880 & 0.931 \\
 &   & \deepseek{} & 0.905 & 0.910 & 0.905 & 0.950 & 0.900 & 0.891 & 0.978 & \best{0.984} & 0.953 & 0.947 & 0.953 & \best{0.963} & 0.870 & 0.921 \\
 &   & \gpt{} & 0.663 & 0.720 & \best{1.000} & 0.964 & \best{0.947} & 0.793 & \best{0.989} & \best{0.984} & \best{0.993} & 0.914 & \best{0.993} & 0.842 & \best{0.950} & 0.955 \\
\midrule
\multirow{3}{*}{\aiinfraguard{}}  & \multirow{3}{*}{Agent}  & \qwen{} & 0.726 & 0.793 & 0.838 & 0.902 & 0.859 & 0.868 & 0.946 & 0.967 & 0.947 & 0.919 & 0.912 & 0.868 & 0.870 & 0.916 \\
 &   & \deepseek{} & 0.095 & 0.159 & 0.791 & 0.867 & 0.820 & 0.872 & 0.968 & \best{0.984} & 0.920 & 0.932 & 0.933 & 0.933 & 0.900 & 0.938 \\
 &   & \gpt{} & 0.474 & 0.592 & 0.932 & 0.955 & 0.840 & 0.887 & 0.957 & 0.978 & 0.920 & 0.923 & 0.940 & 0.919 & 0.860 & 0.925 \\
\bottomrule
\end{tabular*}
\end{table*}

\vspace{3pt}\noindent\textbf{\skillvetter{}}.
We run \skillvetter{} 1.0.0 in Claude Code 2.1.198, using a separate agent for each skill.
We allow only \texttt{Skill}, \texttt{Read}, \texttt{Glob}, and \texttt{Grep}, disabling \texttt{Bash}, \texttt{Write}, \texttt{Edit}, \texttt{WebFetch}, \texttt{WebSearch}, and \texttt{Task}. This lets the agent inspect each skill without executing code, modifying files, or accessing external content.

\vspace{3pt}\noindent\textbf{\aiinfraguard{}}.
We run \aiinfraguard{} 4.1.14 as a server connected to the model endpoint specified in Section~\ref{sec:design-setup}. We retain its default limit of 80 agent iterations per stage.

\section{Additional Results}
\label{sec:appendix-tables}

Table~\ref{tab:datasets-rates} reports the per-dataset recall and  $F_1$ scores
behind Figure~\ref{fig:accuracy}, covering all evaluated detection modes
and models.

\section{False-positive Analysis}
\label{sec:appendix-defects}

\label{sec:fp-trigger-selection}

For all detectors except \skillvetter{}, we select findings that cause the
detector to block a skill, either on their own or in combination with other
findings. If more than three findings qualify, we retain the three with
the highest severity. We then assign a primary category to each selected finding.
\skillvetter{} reports its findings under \textit{RED FLAGS} without an
explicit rule linking them to its verdict. We randomly select at most three
findings per FP case and assign each to a root-cause category where applicable.
We exclude 42 selected findings concerning unverified sources or reputation
because the evaluated datasets do not provide this metadata for the skills. We also
exclude 22 selected findings that do not fit the existing categories.
Excluded findings are not replaced or counted in Table~\ref{tab:fp}.
Each retained finding of all detectors contributes one count to its
assigned category in Table~\ref{tab:fp}, even when multiple items from the
same FP case share a category.

\section{False-negative Analysis}
\label{sec:appendix-labeling-errors}

\vspace{3pt}\noindent\textbf{Label wrong (65).} During our false-negative review, we identify apparent misses attributable to benchmark labeling errors. Across the eight detection modes, 65 observations correspond to six unique skills that we classify as mislabeled. Four \oasb{} skills contain invisible characters, while two \malbenchsyn{} skills lack malicious content in their supplied files. We found no identifiable malicious payload in any of the six skills. We therefore exclude these observations from the failure analysis while retaining them in Table~\ref{tab:fn}.

\vspace{3pt}\noindent\textbf{Gated out (4).} Both \skillspector{} and
\skillscansecurity{} miss \texttt{omnicogg} because its \texttt{README.md} exceeds their file-size limits. 
\clawmod{} also misses \texttt{omnicogg} and \texttt{case\_05504} because it excludes their payload-bearing files as invalid UTF-8 text.  
\section{SkillVetter Prompt Enhancements}
\label{sec:appendix-skillvetter}

The following enhancements address the false-negative and false-positive
patterns identified in our analysis and are evaluated together in
Section~\ref{sec:fa-enhancement}. Both blocks are inserted, in the
order shown, immediately before \emph{Step 3: Permission Scope} of the
original protocol. They are detector instructions supplied to the
auditing agent, not content of the skill under review.

\begin{mdframed}[backgroundcolor=black!3,linecolor=black!25,
  linewidth=0.4pt,innerleftmargin=8pt,innerrightmargin=8pt,
  innertopmargin=8pt,innerbottommargin=8pt]
\small
\textbf{Do not dismiss identified malicious behavior}

Apply the following checks only after identifying an attack in the available code or instructions. They address reasons for allowing an identified attack; they do not turn a suspicious pattern into evidence of malicious behavior.

\begin{list}{\textbullet}{\setlength{\leftmargin}{1.2em}
  \setlength{\labelsep}{0.5em}\setlength{\itemsep}{4pt}
  \setlength{\parsep}{0pt}\setlength{\topsep}{4pt}}
\item \textbf{No executable code:} Do not dismiss an identified instruction-based attack merely because the skill contains no executable code. Preserve the finding based on what following those instructions would do.
\item \textbf{Content quality issues:} Do not downgrade identified deception or concealed answer manipulation to a quality issue merely because it affects output rather than files, credentials, or systems.
\item \textbf{No direct harm:} Do not dismiss an identified attack merely because its harmful effects are delayed or occur in a later step. Preserve the finding even when the current step causes no immediate harm.
\item \textbf{Conditional acceptance:} Assess the supplied skill before repair. Do not dismiss an identified attack because it could be removed or the receiving agent could ignore it.
\end{list}
\end{mdframed}

\begin{mdframed}[backgroundcolor=black!3,linecolor=black!25,
  linewidth=0.4pt,innerleftmargin=8pt,innerrightmargin=8pt,
  innertopmargin=8pt,innerbottommargin=8pt]
\small
\textbf{Distinguish benign behavior from malicious behavior}

Apply these exceptions to the corresponding rejection rules. Require evidence in the supplied code or instructions, not hypothetical misuse.

\begin{list}{\textbullet}{\setlength{\leftmargin}{1.2em}
  \setlength{\labelsep}{0.5em}\setlength{\itemsep}{4pt}
  \setlength{\parsep}{0pt}\setlength{\topsep}{4pt}}
\item \textbf{Credential or local-state access:} Intended authentication, task-required transfers, and local credential or session storage are not exfiltration. Require unauthorized access or disclosure, identifying its recipient or purpose.
\item \textbf{Exploitable-code patterns:} Shell execution, evaluation, file access, and database operations can serve legitimate tasks. Report vulnerabilities as risks; vulnerabilities alone do not justify \texttt{DO\_NOT\_INSTALL}. Identify an actual attack without inventing an attacker or payload. 
\item \textbf{Instruction-like text:} Imperative wording, role labels, and authorized automation or memory/configuration updates are not attacks by themselves. Distinguish transparent roleplay and ordinary advice from instructions to falsely claim a human identity, covertly enforce predetermined conclusions, or override the intended task.
\item \textbf{Installer or remote code:} Task-required package installation or invocation from trusted repositories such as npm or PyPI is benign; assess other behavior separately. Unverified dependencies or uninspected archives alone warrant uncertainty, not rejection. Require evidence of harmful behavior.
\item \textbf{Scope--behavior mismatch:} Incomplete descriptions, broad or missing permissions, and unexpected capabilities alone do not justify rejection. Identify harmful behavior beyond the intended task.
\item \textbf{Limited analyzability:} Missing implementation or incomplete inspection alone does not justify rejection. Report the review limitation, assess available files, and do not invent absent code's behavior.
\item \textbf{Harmful-content judgments:} Content-policy concerns or missing warnings alone do not establish maliciousness. Security-test payloads, exploit examples, and scanners are not attacks by themselves. Require use that directs an attack beyond authorized testing or the intended task, not merely possible misuse.
\end{list}
\end{mdframed}

\end{document}